\documentclass[longauth]{aa}
\usepackage[varg]{txfonts}
\usepackage{graphicx}
\usepackage{natbib}
\usepackage{mathrsfs}
\usepackage{xcolor}
\usepackage{pdflscape}
\usepackage{afterpage}
\usepackage{hyperref}
\usepackage{newtxtext,newtxmath}

\def\deg{\ensuremath{\,{\rm deg}}\xspace}

\def\fh{\ensuremath{^{\mathrm h}}}
\def\fm{\ensuremath{^{\mathrm m}}}
\def\fs{\ensuremath{^{\mathrm s}}}
\def\fdg{\ensuremath{^\circ}}
\def\fmin{\ensuremath{^\prime}}
\def\fsec{\ensuremath{^{\prime\prime}}}
\newcommand\gdr[1]{\gaia~DR#1}

\newcommand{\gaia}{Gaia\xspace}

\newcommand{\exofast}{\texttt{EXOFASTv2}\xspace}
\newcommand{\smw}{\ensuremath{S_{\rm MW}}\xspace}
\def\teff{\ensuremath{T_{\rm eff}}\xspace}
\def\logg{\ensuremath{\log g}\xspace}
\def\gmag{\ensuremath{G}\xspace}
\def\gbp{\ensuremath{G_{\rm BP}}\xspace}
\def\grp{\ensuremath{G_{\rm RP}}\xspace}
\def\parallax{\ensuremath{\varpi}\xspace}
\def\feh{\ensuremath{[\rm Fe/H]}\xspace}
\providecommand{\gcm}{\ensuremath{\,\rm{g}\,\rm{cm}^{-3}}\xspace}
\providecommand{\yr}{\ensuremath{\,\rm yr}\xspace}
\providecommand{\pc}{\ensuremath{\,\rm pc}\xspace}
\providecommand{\Lsun}{\ensuremath{\,L_{\odot}}\xspace}
\providecommand{\Msun}{\ensuremath{\,M_{\odot}}\xspace}
\providecommand{\Rsun}{\ensuremath{\,R_{\odot}}\xspace}
\providecommand{\Mjup}{\ensuremath{\,M_{\rm Jup}}\xspace}
\providecommand{\Mearth}{\ensuremath{\,M_{\oplus}}\xspace}
\providecommand{\mps}{\ensuremath{\,{\rm m\,s}^{-1}}\xspace}
\providecommand{\kmps}{\ensuremath{\,{\rm km\,s}^{-1}}\xspace}
\providecommand{\days}{\ensuremath{\,\rm d}\xspace}
\providecommand{\au}{\ensuremath{\,\rm au}\xspace}
\providecommand{\deg}{\ensuremath{\,\rm deg}\xspace}

\providecommand{\sresinob}{\ensuremath{\sqrt{e_{\rm b}}\sin{\omega_{\rm b}}}\xspace}
\providecommand{\srecosob}{\ensuremath{\sqrt{e_{\rm b}}\cos{\omega_{\rm b}}}\xspace}

\def\mas{\,{mas}\xspace}
\def\masyr{{mas yr$^{-1}$}\xspace}

\bibpunct{(}{)}{;}{a}{}{,}
\graphicspath{{images/}}

\title{The GAPS programme at TNG}
\subtitle{LXXV. Validating and confirming Gaia substellar astrometric candidates with HARPS-N}

\author{D.~Barbato          \inst{\ref{oapd}}
        \and M.~Pinamonti   \inst{\ref{oato}}
        \and A.~Sozzetti    \inst{\ref{oato}}
        \and S.~Desidera    \inst{\ref{oapd}}
        \and V.~D'Orazi     \inst{\ref{uniroma},\ref{oapd}}
        \and J.~Maldonado   \inst{\ref{oapa}}
        \and K.~Biazzo      \inst{\ref{oar}}
        \and L.~Naponiello  \inst{\ref{oato}}
        \and A.~F.~Lanza    \inst{\ref{oact}}
        \and A.~Bignamini   \inst{\ref{oats}}
        \and A.~S.~Bonomo   \inst{\ref{oato}}
        \and M.~Brogi       \inst{\ref{oato},\ref{unito}}
        \and L.~Cabona      \inst{\ref{oabr}}
        \and M.~Damasso     \inst{\ref{oato}}
        \and R.~Gratton     \inst{\ref{oapd}}
        \and L.~Mancini     \inst{\ref{uniroma},\ref{oato},\ref{mpia}}
        \and G.~Mantovan    \inst{\ref{cisas},\ref{oapd}}
        \and D.~Nardiello   \inst{\ref{unipd},\ref{oapd}}
        \and M.~Rainer      \inst{\ref{oabr}}
        \and G.~Guilluy     \inst{\ref{oato}}
        \and P.~Giacobbe    \inst{\ref{oato}}
        \and L.~Malavolta   \inst{\ref{oapd},\ref{unipd}}
        \and R.~Cosentino   \inst{\ref{fgg}}
        \and W.~Boschin     \inst{\ref{fgg},\ref{iac},\ref{unilaguna}}
        \and R.~Claudi      \inst{\ref{oapd},\ref{uniroma3}}
        }
\institute{
            INAF – Osservatorio Astronomico di Padova, Vicolo dell’Osservatorio 5, I-35122, Padova, Italy \\ \email{domenico.barbato@inaf.it} \label{oapd}
            \and INAF – Osservatorio Astrofisico di Torino, Via Osservatorio 20, I-10025, Pino Torinese, Italy \label{oato}
            \and Department of Physics, University of Rome Tor Vergata, via della Ricerca Scientifica 1, I-00133, Rome, Italy \label{uniroma}
            \and INAF – Osservatorio Astronomico di Palermo, Piazza del Parlamento 1, I-90134 Palermo, Italy \label{oapa}
            \and INAF – Osservatorio Astronomico di Roma, Via Frascati 33, I-00078, Monte Porzio Catone (Roma), Italy \label{oar}
            \and INAF – Osservatorio Astrofisico di Catania, Via S. Sofia 78, I-95123 Catania, Italy \label{oact}
            \and INAF – Osservatorio Astronomico di Trieste, via Tiepolo 11, I-34143 Trieste \label{oats}
            \and Dipartimento di Fisica, Universit\`a degli Studi di Torino, via P. Giuria 1, Turin, I-10125, Italy \label{unito}
            \and INAF – Osservatorio Astronomico di Brera, Via E. Bianchi 46, I-23807 Merate, Italy \label{oabr}
            \and Max Planck Institute for Astronomy, Königstuhl 17, I-69117 Heidelberg, Germany \label{mpia}
            \and Dipartimento di Fisica e Astronomia Galileo Galilei, Universit\`a degli Studi di Padova, Vicolo dell’Osservatorio 3, I-35122 Padova, Italy \label{unipd}
            \and Centro di Ateneo di Studi e Attività Spaziali G. Colombo – Universit\`a degli Studi di Padova, Via Venezia 15, I-35131, Padova, Italy \label{cisas}
            \and Fundacion Galileo Galilei – INAF, Rambla J.A. Fernandez P., 7, E-38712 S.C.Tenerife, Spain \label{fgg}
            \and Instituto de Astrofísica de Canarias, C/Vía Láctea s/n, 38205 La Laguna (Tenerife), Canary Islands, Spain \label{iac}
            \and Departamento de Astrofísica, Univ. de La Laguna, Av. del Astrofísico Francisco Sánchez s/n, 38205 La Laguna (Tenerife), Canary Islands, Spain \label{unilaguna}
            \and Dipartimento di Matematica e Fisica, Universit\`a Roma Tre, Via della Vasca Navale 84, 00146 Roma, Italy \label{uniroma3}
            }

\date{Received 7 October 2025 / Accepted 19 June 2026}
\abstract{The astrometric measurements provided by the Gaia space mission represent a key advancement in the search and characterization of exoplanets, helping in particular to solve the mass degeneracy intrinsic to the radial velocity (RV) method. The fact that a fraction of astrophysical false positives contaminates the current catalog of astrometric candidate solutions requires an RV follow-up to validate and confirm such candidates.}
{Within the GAPS programme, we have observed a selected sample of 14 stars having Gaia astrometric solutions compatible with the presence of a substellar companion. The immediate aim of this survey is to identify astrophysical false positives and provide the first RV validation and confirmation of the remaining candidates.}
{We analysed data collected with the HARPS-N spectrograph to identify stellar binary systems from the spectral cross-correlation function profiles. The remaining astrometric candidates were characterized via Markov chain Monte Carlo analysis searching for the best-fit RV solution.}
{Among the stars in our sample with astrometric candidate solutions, we identify 6 as originating from close binary companions mimicking the astrometric motion of distant substellar companions, from which we can estimate an updated value of $43_{-11}^{+13}$\% for the binary contamination fraction in the Gaia DR3 catalog of astrometric candidates. We validate and confirm the remaining 8 solutions, corresponding to giant and brown dwarf companions with minimum masses between 8 and 62 $M_{\rm Jup}$ and semimajor axes between 0.76 and 1.42 au, providing the first RV characterization for 6 of these candidates and updated orbital solutions for 2 previously confirmed ones.}
{}

\begin{document}
    \maketitle
  
    \section{Introduction}  \label{sec:introduction}
        A key issue in current exoplanetology is the determination of the occurrence rate of systems in which companion mass tends to increase with orbital separation. Such so-called ordered systems \citep{mishra2023} are of special interest in investigating the origin of the observed variety of system architectures, with theoretical models advocating various possible formation pathways and occurrence rates. Current models argue for a weak \citep[e.g][]{schlecker2021}, positive \citep[e.g.][]{bitsch2023} or negative \citep[e.g][]{izidoro2015a,morbidelli2015} effect of the presence of cold Jupiters ($a>1$\au, $0.3<M<20$\Mjup) on the formation and evolution of inner super-Earths ($2<M<10$\Mearth) and Neptunes ($10<M<20$\Mearth).
        \par Over the years, many observational campaigns have attempted to measure the occurrence rates of such ordered systems and investigate their origins. Some works find strong positive correlations between outer giants and inner low-mass exoplanets \citep{zhu2018,bryan2019,bryan2024,lefevreforjan2025,bonomo2025}, while others report either a weak correlation \citep{rosenthal2022}, no correlation \citep{bonomo2023} or a negative correlation \citep{barbato2018}. Although a significant part of such a tension between observational results arises from non-trivial comparisons between different definitions of planetary categories, system multiplicity and stellar properties in the analysed samples, another fundamental issue arises from the fact that the vast majority of intermediate separation ($1<a<5$\au) cold Jupiters has been detected by radial velocity (RV) observations alone and are therefore characterized by only its minimum mass $M\sin{i}$. Additionally, many studies focusing on determining the occurrence rate of low-mass companions in the presence of cold Jupiters \citep[e.g.][]{bryan2019,rosenthal2022} often infer the outer companion presence from long-term RV trends, which are intrinsically degenerate with mass and separation. Therefore, RV-only studies of the demography of ordered systems suffer from the limitations imposed by the incomplete characterization of the outer companions' orbital characteristics and true dynamical mass.
        \par Similarly, the true mass of brown dwarfs (BD) represents a key input for formation and evolution models for these companions. While BDs are generally thought to be formed via gravitational instability \citep{chabrier2014,whitworth2018}, significant work is still needed to investigate the origin of the observed scarcity of BDs on close orbits around solar-type stars (the so-called brown dwarf desert), with different evolution pathways being proposed \citep[see e.g.][]{armitage2002,jumper2013,ma2014,maldonado2017bd}. Solving this issue requires large statistics and reliable true mass measurements, and as such RV-only exploration of the BD population suffers from the same limitations discussed for giant exoplanets.
        \par ESA’s Gaia space astrometry mission \citep{prusti2016} represents an essential turning point, having the potential to detect thousands of cold Jupiters and BDs as distant as 5 au from their host stars \citep{casertano2008,sozzetti2014,perryman2014}. Furthermore, the high-precision proper motion measurements provided by Gaia have enabled the analysis of proper motion anomaly (PMa) between the Hipparcos and Gaia epochs \citep[e.g.][]{brandt2021hgca,kervella2022}. The third Gaia Data Release \citep[\gdr{3},][]{vallenari2023} has produced a catalog of 169227 non-single stars (NSS) for which the observed astrometric motion solution is consistent with a Keplerian orbit produced by an unseen massive companion. Of particular interest amongst these are $\sim$2000 orbits consistent with companions having estimated true dynamical masses in the giant planet and BD regime. These candidates represent a fertile ground for follow-up observations resulting in the validation and characterization of cold Jupiters such as HIP\,66074\,b \citep[Gaia-3\,b,][]{winn2022,marcussen2023,sozzetti2023}, Gaia-4\,b \citep{stefansson2025}, as well as BDs like LHS\,1610\,b \citep{fitzmaurice2024} and Gaia-5\,b \citep{stefansson2025}. This continuous synergy between RVs and astrometry, additionally fuelled by future Gaia DRs, is a key asset in solving RV mass degeneracy and will allow the firm assessment of ordered system demographics based on true masses and orbits of outer companions \citep[see e.g.][]{sozzetti2024}.
        \par While encouraging, the \gdr{3} exoplanetary potential is still limited by the presence of astrophysical false positives. Indeed, the observed photocenter motion of a binary system can be as small as the one caused by a substellar companion which would instead contribute no flux, provided that the flux ratio of the stellar binary components is similar to their mass ratio \citep[see e.g.][for further details]{arenou2023,marcussen2023}. As such, it is known that a portion of the small photocenter motions observed by Gaia can be caused by a stellar binary companion mimicking the astrometric motion produced by a massive substellar companion. \cite{holl2023} estimated that $\sim10$\% of DR3 astrometric orbital solutions consistent with substellar companions may actually be caused by this binary contamination effect. Luckily, such astrophysical false positives can be recognized by RV follow-up campaigns, as high signal-to-noise ratio (S/N) spectra can identify double-lined spectroscopic binaries (SB2). Indeed, the RV follow-up campaign of 28 NSS candidates detailed in \cite{stefansson2025} results in an updated estimate of binary contamination of $\sim$32\%, a value notably larger than the one derived in \cite{holl2023}. As such, RV follow-up campaigns of promising Gaia NSS compatible with planetary or BD companions represent a key asset in recognizing false positives and provide an updated assessment of the binary contamination in DR3 astrometric solutions in preparation for future Gaia releases.
        \par In this context, the GAPS \citep[Global Architecture of Planetary Systems, see][]{covino2013,desidera2013} collaboration\footnote{\url{https://www.gaps.inaf.it}} started in October 2023 an intense RV observational campaign of 35 stars orbited by known or confirmed cold Jupiters and low-mass brown dwarfs using the High Accuracy Radial velocity Planet Searcher in the Northern hemisphere \citep[HARPS-N, see][]{cosentino2012} spectrograph mounted on the Telescopio Nazionale Galileo (TNG) in La Palma. Of these outer companions, 21 have true masses and full orbits measured based on the combination of literature RVs, PMa and \gdr{3} astrometric orbital solutions, while the remaining 14 are Gaia NSS candidates in need of validation through RV follow-up observations lacking any public RV data at the moment of their inclusion in our sample. These 14 targets have been selected from NSS catalog by virtue of their distance ($<$200\pc), brightness (\gmag$<$12\,mag), declination ($\delta>-10$\,deg) and for having astrometric candidate solutions compatible with an estimated true companion mass $<$65\Mjup and orbital separation between 1 and 2\au. The aim of this ongoing observational campaign is three-fold: identify astrophysical false positives in the selected \gdr{3} astrometric orbital solutions, provide first RV validation of the remaining candidates and, finally, perform a deep search for short-period super-Earths and Neptunes in the presence of known massive outer companions to measure directly the occurrence rate of ordered systems.
        \par In this paper we present the first results of this ongoing campaign, focusing on the refutation and validation of 14 Gaia NSS astrometric candidates, while additional noteworthy results from this campaign will be presented in \cite{pinamonti2026} and other forthcoming papers. This work is structured as follows: in Sect.~\ref{sec:star-parameters} we provide an updated description of the physical characteristics of the stars populating the sample analysed in this work, before focusing in Sect.~\ref{sec:binaries} on the identification of 6 astrophysical false positives contaminating the astrometric solutions. In Sect.~\ref{sec:validation}, we turn our attention towards the confirmation of 8 astrometric solutions through the robust characterization of full RV orbits. We finally discuss our results and the future perspectives of the ongoing Gaia follow-up survey in Sect.~\ref{sec:conclusions}.
        
    \section{Stellar parameters}  \label{sec:star-parameters}
        \par We derived spectroscopic stellar parameters for the stars in our sample using the HARPS-N spectra collected during our survey, following different procedures depending on the star spectral type. We note that this approach is not applicable to the stars we identify as SB2 (see Sect.~\ref{sec:binaries}) due to the intrinsic difficulties in disentangling the spectra of the two stellar components. As a result, the following spectroscopic analysis refers to the stars hosting validated companions (see Sect.~\ref{sec:validation}.
        \par For stars with spectral types up from late-K/early-M we compute basic stellar parameters (effective temperature, surface gravity, iron abundance) using a methodology developed by \cite{maldonado2015} and based on ratios of spectral features \citep[see also ][]{maldonado2017}. For stars with spectral types up to mid-K we instead follow the procedure described in \cite{biazzo2022}. In particular, we considered the list of iron lines by the latter authors and \cite{dalponte2025} and we then measured the line equivalent widths with the \texttt{ARESv2} tool \citep{sousa2015}, then double-checked with \texttt{IRAF} \citep{tody1993}. We considered the \texttt{MARCS} atmospheric models \citep{gustafsson2008} and applied the \texttt{pyMOOGi} code \citep{sneden1973,adamow2017} as spectral analysis package. We thus derived final effective temperature T$_{\rm eff}$, surface gravity $\log{g}$, iron abundance \feh and microturbulence velocity v$_{\rm micro}$ (listed in Table~\ref{table:star-validated}).
        \par In order to have an updated characterisation of stellar mass, radius, density, luminosity and age we fit the stellar spectral energy distributions (SED) using the MESA Isochrones and Stellar Tracks (MIST) \citep{dotter2016,choi2016} via the \texttt{IDL} suite \exofast \citep[see][for a detailed description of the method]{eastman2019} considering all available archival magnitudes. We impose Gaussian priors on the star's parallax \parallax based on \gaia~DR3 astrometric measurement, as well as on effective temperature \teff and iron abundance \feh based on their values obtained as detailed in the previous paragraph. For the stars we identify as SB2, for which the spectroscopic analysis was not possible, we base the \teff and \feh priors on the respective values in the \cite{anders2022} photo-astrometric catalog. We neglect the effects of extinction, as most of the stars in our sample are relatively nearby ($<$150\pc) and therefore not affected by significant extinction, while the most distant stars in our sample are identified as SB2 in Section~\ref{sec:binaries}.
        \par We additionally compute the kinematic ages for the stars in our sample following \cite{almeida2018}, computing the Galactic velocity vector (U, V, W) from each star's systematic radial velocity, proper motions and parallax. When available, we use the 25 yr average proper motions listed in the Hipparcos-Gaia Catalog of Accelerations \citep[HGCA, see][]{brandt2021hgca}, using the \gdr{3} values for the stars not in the HGCA.
        \par We additionally use the Python tool \texttt{ACTIN2}\footnote{\url{https://actin2.readthedocs.io/en/latest/index.html}} \citep{gomesdasilva2018, gomesdasilva2021} to extract activity indicators from the collected HARPS-N spectra, considering in particular emission from the Ca II H\&K (\smw), H$\alpha$, Na I D2 and He I lines. We note that no stars in our sample exhibit strong activity levels, and we analyse the activity indices time series and their relationships with the observed RV variations for each star in Section~\ref{sec:validation}.
       \par All relevant archival stellar information, as well as the parameters resulting from the analyses detailed in this Section, are listed in Table \ref{table:star-binary} for the SB2 stars and in Table~\ref{table:star-validated} for the stars hosting Gaia candidates we validate in this work. For the latter, we note that our derived stellar masses are compatible within 1.5$\sigma$ with the adopted primary masses in the NSS catalog, which we report for each star in Section~\ref{sec:validation}. We must note, however, that the SED fit performed for the SB2s may be influenced by their binary nature and lead to less-than-robust results.

    \section{Binary contamination}  \label{sec:binaries}
        \begin{figure}
            \includegraphics[width=\linewidth]{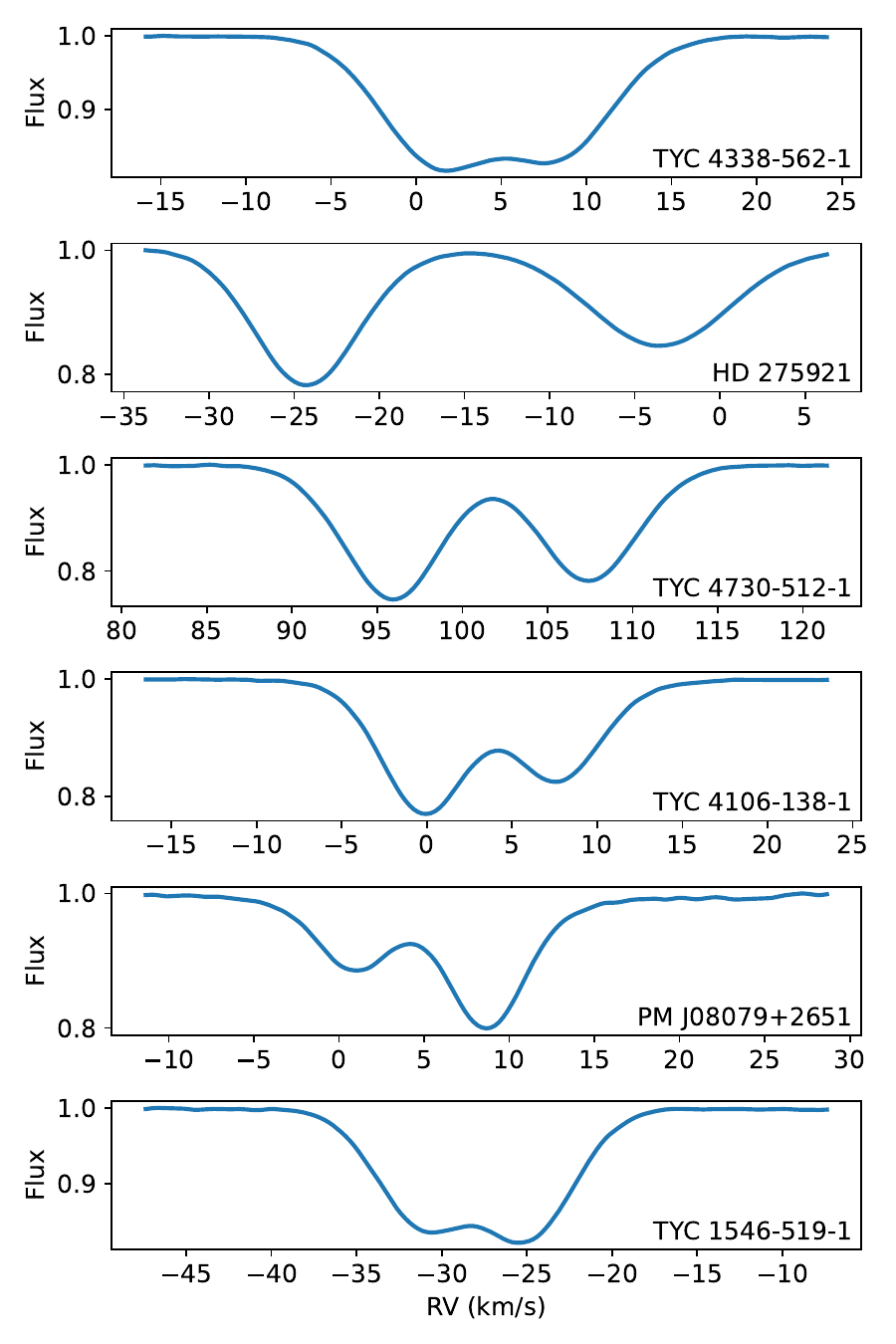}
            \caption{HARPS-N cross correlation function profiles of the 6 stars in the sample identified as binaries mimicking substellar astrometric motion.}
            \label{fig:binary-ccf}
        \end{figure}
        As mentioned in Sect.~\ref{sec:introduction}, stellar binaries are noted to be the main source of such false positives \citep{arenou2023,holl2023,marcussen2023}, with binary components having a mass ratio similar to the flux ratio being notably able to mimic the small astrometric motions caused by substellar companions. As such, during our RV observations with HARPS-N special care was taken to produce and analyse the cross-correlation functions (CCF) profiles of the high-S/N spectra (ranging from 40 to 70), reduced using the HARPS-N data reduction software \citep[DRS,][]{lovis2007}, to identify the double-peaked profiles of SB2 configurations and exclude such stars from subsequent observations.
        \par Amongst the 14 stars having \gdr{3} astrometric candidate solutions included in our sample, we identify 6 stars as being SB2 systems generating astrometric false positives solutions, for which we show in Fig.~\ref{fig:binary-ccf} the HARPS-N CCF having a clear double-peak profile. In Table~\ref{table:star-binary} we also report the candidate Gaia non-single star orbital period $P_{\rm NSS}$. We note that two of these stars, PM\,J08079+2651 and TYC\,1546-519-1, have a \gdr{3} SB1 solution.
        \par Having found 6 astrophysical false positives amongst the 14 \gdr{3} candidate solutions, we can derive a $43_{-11}^{+13}$\% binary contamination fraction assuming binomial uncertainties, a value that is notably higher than the $\sim$10\% estimated in \cite{holl2023} but instead fully compatible within 1$\sigma$ with the more recent $\sim$32\% derived by \cite{stefansson2025}.

    \section{Gaia candidates validation and confirmation} \label{sec:validation}
        \begin{table*}
            \caption{HARPS-N measurements for the stars discussed in this work.}    \label{table:rvdata}
            \centering
            \tiny
            \begin{tabular}{l c c c c c c c c c}
                \hline\hline
                    Star & BJD & T$_{\rm exp}$ & RV & CCF FWHM & CCF BIS & \smw & H$\alpha$ & Na~{\sc i}\\
                        & & (s) & (\kmps) & (\kmps) & (\kmps)  & & & \\[3pt]
                \hline
                    2MASS J20570943+1406590   & $2460224.50$ & 1800 & $6.7426\pm0.0059$ & $5.9003$ & $0.0358$ & $0.1662\pm0.0153$& $0.1547\pm0.0023$& $0.3018\pm0.0044$\\
                    2MASS J20570943+1406590   & $2460227.54$ & 1800 & $6.7014\pm0.0030$ & $5.8662$ & $0.0089$ & $0.2156\pm0.0083$& $0.1541\pm0.0015$& $0.2982\pm0.0029$\\
                    2MASS J20570943+1406590   & $2460236.46$ & 1800 & $6.5794\pm0.0023$ & $5.8740$ & $0.0132$ & $0.1469\pm0.0066$& $0.1552\pm0.0014$& $0.2954\pm0.0027$\\
                    2MASS J20570943+1406590   & $2460244.45$ & 1800 & $6.4603\pm0.0024$ & $5.8894$ & $0.0322$ & $0.1520\pm0.0057$& $0.1575\pm0.0015$& $0.2968\pm0.0027$\\
                    2MASS J20570943+1406590   & $2460245.45$ & 1800 & $6.4460\pm0.0027$ & $5.8817$ & $0.0129$ & $0.1476\pm0.0061$& $0.1565\pm0.0016$& $0.3041\pm0.0029$\\
                    ... & ... & ... & ... & ... & ... & ... & ... & ...\\
                \hline
            \end{tabular}
            \tablefoot{Data are available at the CDS. A portion is shown here for guidance regarding its form and content.}
        \end{table*}
        Having identified and excluded 6 new false positives among the 14 Gaia astrometric candidate orbits in our sample, we focus in this Section on the remaining 8 substellar Gaia astrometric candidates validated by single-peaked CCF profile. As mentioned in Sect.~\ref{sec:binaries}, the spectra collected were reduced using the HARPS-N DRS \citep{lovis2007}, while the RV extraction was performed using the Template-Enhanced Radial velocity Re-analysis Application pipeline \citep[TERRA,][]{angladaescude2012}. All RV measurements collected, as well as all activity indicators analysed in the following, are listed in Table~\ref{table:rvdata}, while in Table~\ref{table:rvfit} we list the number of the measurements collected, the observational timespan, and the coverage of the NSS orbital period, average RV uncertainty and weighted root mean square (w.r.m.s).
        \par To search for massive long-period companions in the HARPS-N RV time series we collected during our observations, we calculated the generalized Lomb-Scargle periodogram \citep[GLS,][]{zechmeister2009} of the RVs to identify significant signals having a bootstrap False Alarm Probability (FAP) lower than 0.1\%. At the same time, we analyse the activity indices time series derived as discussed in Sect.~\ref{sec:star-parameters} to investigate any stellar activity origin for the observed significant RV signals. Similarly, we inspect the available All Sky Automated Survey \citep[ASAS, ][]{pojimanski2002}, Transiting Exoplanet Survey Satellite \citep[TESS, ][]{ricker2015} and Gaia light curves for each star in the sample to identify via GLS and autocorrelation function \citep{mcquillan2013} any modulation due to stellar activity compatible with the timeline of the photometric observations. When appropriate, we also model the available light curves with Gaussian Process (GP) and quasi-periodic kernel fits to derive stellar rotational period P$_{\rm rot}$, in order to inform future searches for close-in companions. Unless otherwise stated, our HARPS-N observations represent the first RV time series available for these stars, as no other public RV measurements are found in the literature or in the archives.
        \par
\begin{figure*}
            \centering
            \includegraphics[width=0.497\linewidth]{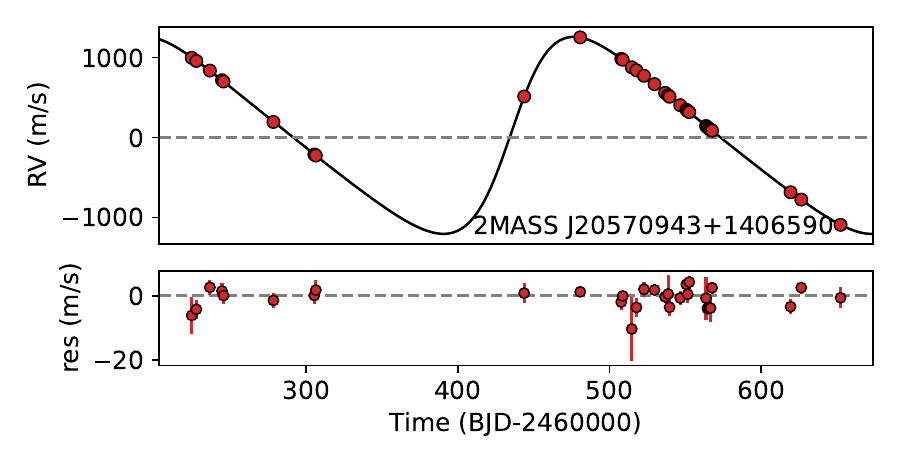}
            \includegraphics[width=0.497\linewidth]{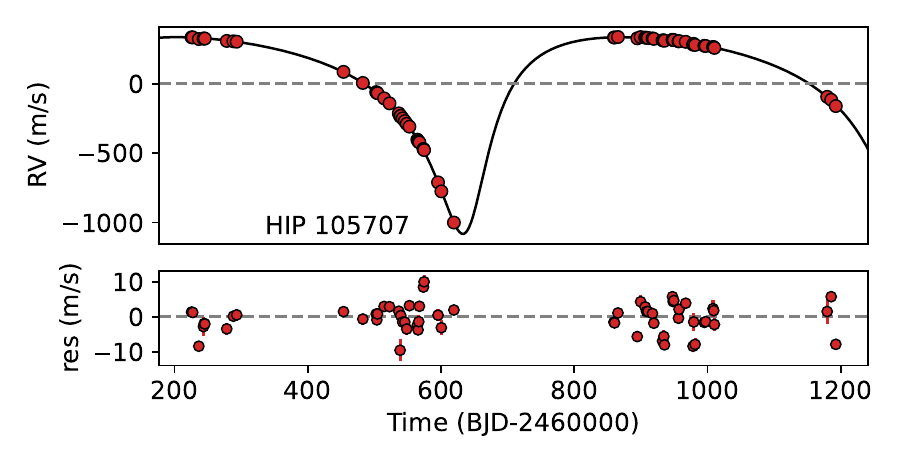}
            \includegraphics[width=0.497\linewidth]{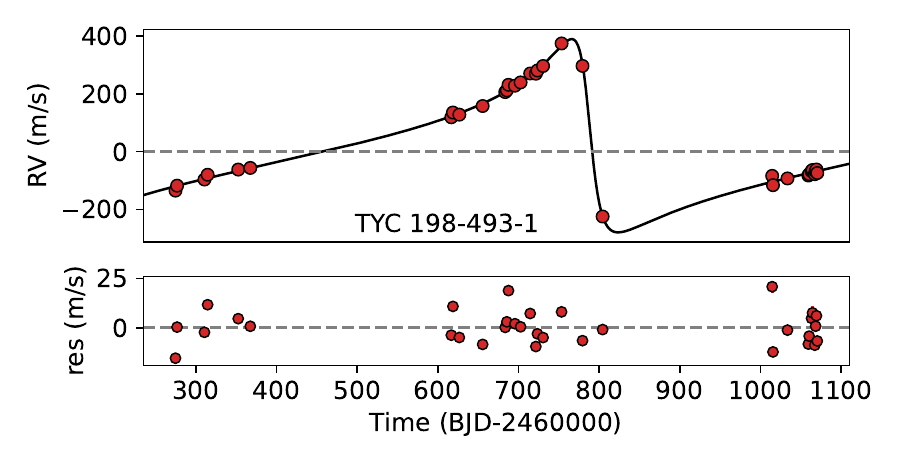}
            \includegraphics[width=0.497\linewidth]{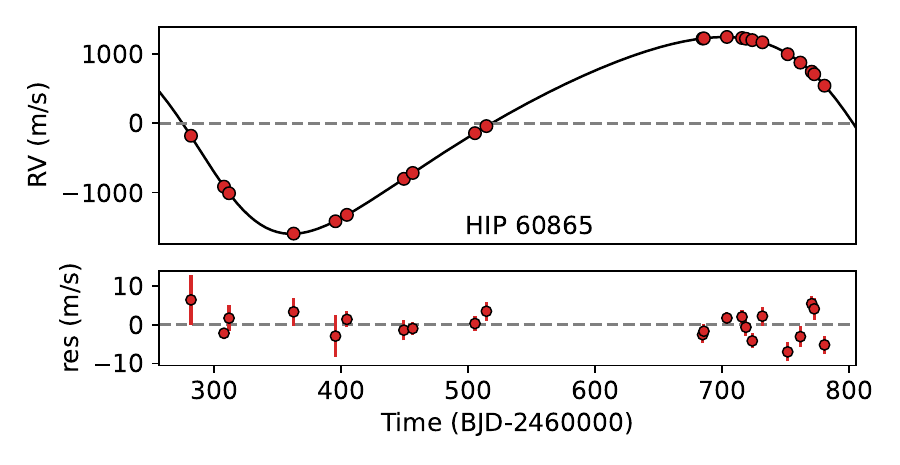}
            \includegraphics[width=0.497\linewidth]{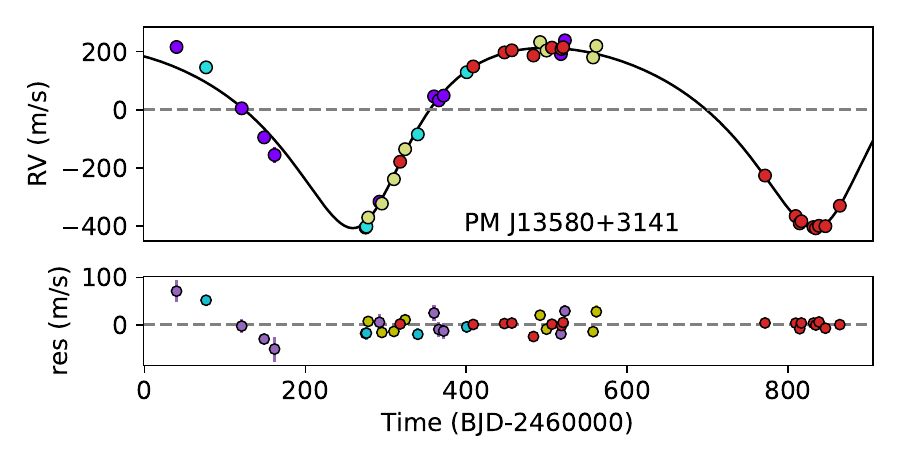}
            \includegraphics[width=0.497\linewidth]{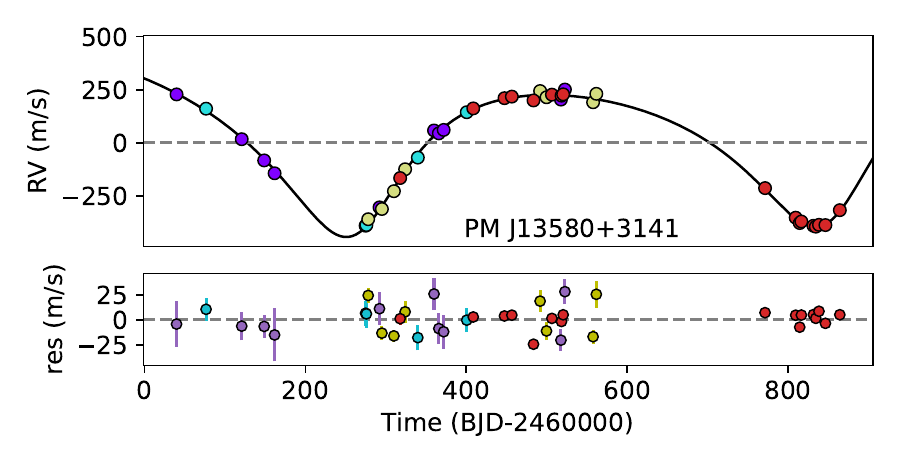}
            \includegraphics[width=0.497\linewidth]{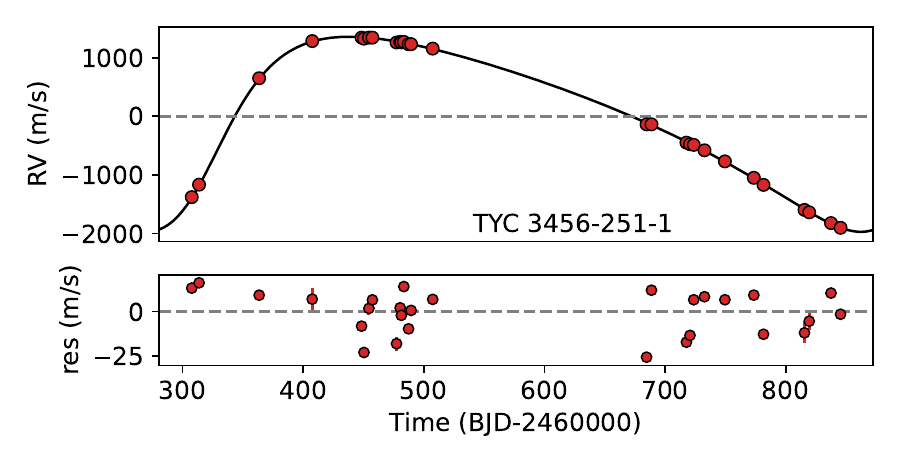}
            \includegraphics[width=0.497\linewidth]{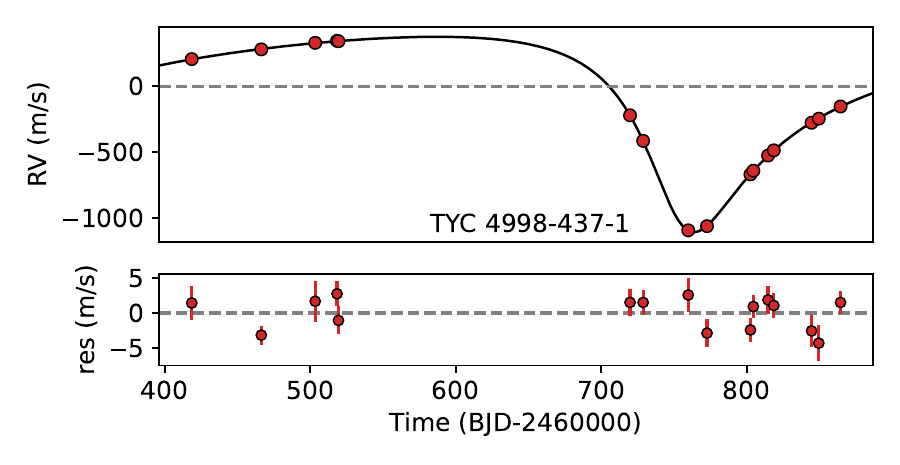}
            \includegraphics[width=0.497\linewidth]{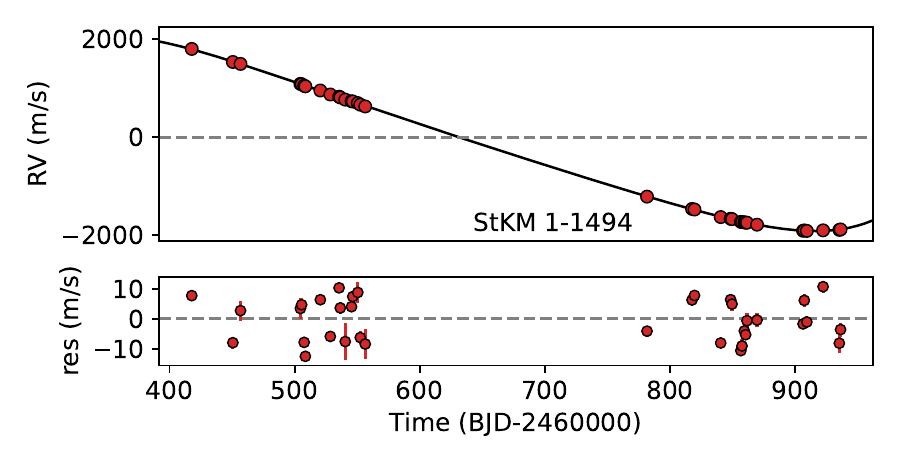}
            \caption{Radial velocity orbital fits for the stars discussed in this work. For all stars, The top panel shows the best-fit solution over the HARPS-N (red circles) data, while the bottom panel shows the post-Keplerian fit residuals. In the two figures shown for PM\,J13580+3141 (single-Keplerian and Keplerian with quadratic trend solutions respectively), literature HPF, FIES and NEID data are shown as purple, cyan and yellow circles.}
            \label{fig:rvfit}
        \end{figure*}

 For each star for which we identify significant and robust RV signals, we search for the best-fit orbital solution using the Python tool \texttt{PyORBIT}\footnote{\url{https://github.com/LucaMalavolta/PyORBIT}} \citep{malavolta2016,malavolta2018}, a package for the Markov chain Monte Carlo (MCMC) modelling of RV and activity indices time series based on the optimization algorithm \texttt{PyDE} \citep{storn1997} and the MCMC sampler \texttt{emcee} \citep{foreman2013}. The free parameters we fit for are orbital period $P$, semi-amplitude $K$, mean longitude $\lambda_0$, \sresinob, \srecosob, and finally a zero-point radial velocity term $\gamma$ and an uncorrelated stellar jitter term $j$ for each RV dataset. We set a uniform prior on $P$ based on the astrometric candidate period nominal value $P_{\rm NSS}$, bounding the fitted $P$ between 0.8 and 1.2 times this value. We detail the results of our analysis for each star in the following subsections, presented in decreasing order of HARPS-N coverage of the astrometric candidate orbital period, showing the orbital solutions in Fig.~\ref{fig:rvfit} and listing the best-fit parameters in Table~\ref{table:rvfit}. In the same Table we additionally report the relevant NSS candidate solution orbital parameters, as well as the companion mass lower value $M_{\rm NSS}$ corresponding to the lower confidence level assuming no flux contamination, and true mass estimates $M_{b,i_{\rm NSS}}$ from our RV minimum mass values by adopting the NSS orbital inclination value to provide additional comparison between our RV-only solution and the NSS candidate parameters. Unless otherwise stated in the following subsections, we generally find our RV-only bestfit parameters to be within 2$\sigma$ of the NSS candidate solution values.
        
        \subsection{2MASS\,J20570943+1406590}   \label{subsec:gfu24}
            \par 2MASS\,J20570943+1406590 is a K4 star located at 110\pc from the Sun. No long-period modulation is evident from the available ASAS and TESS (Sectors 55 and 82) light curves. The \gdr{3} NSS solution is compatible with the presence of a companion having an orbital period of $280.76\pm3.16$\days, eccentricity of $0.33\pm0.13$ and, assuming a 0.72\Msun mass for the host star, a lower 1-$\sigma$ confidence level on true mass (hereinafter referred to as lower $M_{\rm NSS}$ for simplicity) of $\sim$38.78\Mjup \citep{gaiabinmass2022}.
            \par We show in the top row of Fig.~\ref{fig:GFU24-timeseries} the HARPS-N RV time series and GLS periodogram, from which we find a highly significant principal peak at $\sim$285\days having FAP=$4\cdot10^{-17}$\%. We find no correlation between this RV signal, the CCF full width at half maximum (FWHM) and CCF bisector inverse span (BIS) (shown in the third and fourth rows of Fig.~\ref{fig:GFU24-timeseries}) and any of the activity indices we derived with \texttt{ACTIN2} (see Sect.~\ref{sec:star-parameters}) and shown in the fifth-to-eighth rows of Fig.~\ref{fig:GFU24-timeseries}. In fact, the strongest correlation we find is the one between the RV time series and Na I, having a Pearson correlation coefficient $\rho$=0.43 with a p-value of 0.01 indicating only moderate correlation, with the activity index GLS periodogram having a FAP=0.05\% principal peak located at $\sim$356\days. We therefore exclude a stellar activity origin for the large amplitude RV signal.
            \par Searching for the best-fit orbital solution using \texttt{PyORBIT} we find a single-Keplerian orbit with period of $281.89\pm0.09$\days, semi-amplitude of $1233.31_{-3.91}^{+3.99}$\mps and eccentricity of $0.313\pm0.002$. Using the host star mass value of $0.689^{+0.031}_{-0.030}$\Msun we obtained from the SED fit detailed in Sect.~\ref{sec:star-parameters}, we derive a companion minimum mass of $30.32_{-0.91}^{+0.84}$\Mjup and an orbital semi-major axis of $0.75\pm0.01$\au. The post-Keplerian fit residual RVs GLS periodogram, shown in the second row of Fig.~\ref{fig:GFU24-timeseries}, exhibits no significant signal, with the residual principal peak at $\sim$24\days having a 10\% FAP, nor any strong correlation with the analysed activity indices.
            \par Notably, our RV best-fit solution is highly compatible with the \gdr{3} astrometric candidate. We therefore validate the \gdr{3} solution and assess the observed astrometric motion of the star as caused by the presence of the low-mass brown dwarf companion described in this Section.
            
        \subsection{HIP\,105707} \label{subsec:gfu25}
            HIP\,105707 is a K7 star found at 50\pc away from the Sun. Our analysis of the ASAS and TESS (Sectors 15, 55, and 56) light curves reveals no significant long-term activity. However, a modulation is detected in the TESS data on a timescale of either $\sim$4.4 days (from a single-term Lomb-Scargle and GP fit) or $\sim$8.8 days (from a multi-term Lomb-Scargle fit), suggesting activity on shorter timescales. The NSS solution for this star finds a candidate companion on a $687.71\pm12.67$\days orbit with an eccentricity of $0.51\pm0.03$ and lower $M_{\rm NSS}\sim$23.77\Mjup assuming a primary mass of 0.70\Msun.
            \par Our HARPS-N observations
            (see Fig.~\ref{fig:GFU25-timeseries}) exhibit clear long-term eccentric Keplerian-like behaviour, with a principal periodogram peak of $\sim$604\days with a low FAP value of $1.18\cdot10^{-23}$\%. Most of the activity indices show low-to-moderate correlation with the RVs, with the only exception of high correlation present ($\rho=-0.88$, p-value$=0$) with the H$\alpha$ index. However, the observed RV variation ($\sim$1335\mps) is incompatible with an activity-only origin and is corroborated by the astrometric candidate solution, which is mostly insensitive to such activity signals \citep[see e.g.][]{lagrange2011,meunier2020}, having an orbital period of the same order of magnitude of the observed RV signal, and a companion-induced origin for the RV variation is preferred.
            \par Our Keplerian solution is highly compatible with the astrometric candidate solution, having an orbital period of $668.76_{-2.60}^{+2.54}$\days, semi-amplitude of $707.60_{-7.15}^{+7.78}$\mps and eccentricity $0.563\pm0.005$. From the $0.707^{+0.028}_{-0.026}$\Msun host star mass we derived from our SED fit, we infer a companion minimum mass of $20.34\pm0.58$\Mjup and a semi-major axis of $1.34\pm0.02$\au.
            \par We additionally note that HIP\,105707 is one of the two stars among those discussed in this work for which PMa value is present in the HGCA. As such, we perform a combined fit of the HARPS-N RV observations and PMa measurements using the Python code \texttt{orvara} \citep{brandt2021orvara}. We find for the detected companion a semi-major axis of $1.35\pm0.02$\au, an eccentricity of $0.563_{-0.005}^{+0.006}$, an orbital inclination of ${100}_{-17}^{+16}$\deg which we note to be in agreement with the NSS value within 1.5$\sigma$ and a dynamical mass of ${21.13}_{-0.94}^{+1.80}$\Mjup. This joint RV-PMa solution is in notable agreement with the RV-only one, with the added benefit of providing a true mass value confirming the brown dwarf nature of the detected companion.
            
        \subsection{TYC\,198-493-1} \label{subsec:gfu11}
            TYC\,198-493-1 is a K0 star at 78\pc from the Sun. Although neither the ASAS not TESS (Sectors 07, 34 and 61) data exhibit long-term modulation, they show clear short-term modulation suggesting moderate activity, and we derive a rotational period of $12\pm4$days and $12.9\pm0.4$\days respectively. The NSS solution for this star suggests the presence of a companion with an orbital period of $722.60\pm8.79$\days and high $0.75\pm0.07$ eccentricity and, assuming a host star mass of 0.83\Msun, lower $M_{\rm NSS}\sim$49\Mjup.
            \par As for HIP\,60865, four archival SOPHIE RV measurements are available, but their $\sim$0.1\kmps uncertainty renders them usuitable for our analysis. As shown in Fig.~\ref{fig:GFU11-timeseries}, a large long-term Keplerian variation is present in the data. Apart from a strong 1\days peak in the periodogram, the principal peak is located at $\sim$658\days and with a FAP of 0.18\%. No strong correlation between RVs and activity indices is evident, with the exception of a moderate correlation ($\left|\rho\right|\sim$0.58 with p-value $<0.001$) being present between RVs and Na I. We find the data to be best described by a Keplerian model having an orbital period of $718.67_{-6.35}^{+6.41}$\days, semi-amplitude $335.09_{-3.89}^{+4.21}$ and high eccentricity of $0.739\pm0.006$. This solution, given the $0.817^{+0.037}_{-0.035}$\Msun host star mass obtained from our SED fit, corresponds to a $8.63\pm2.52$\Mjup minimum mass and $1.47_{-0.23}^{+0.19}$\au semi-major axis for the companion.
            \par We note that our value of $M\sin{i}$ is much lower than the minimum NSS true mass estimate of 49\Mjup. However, assuming the NSS orbital inclination value of $14.62\pm18.02$\deg our $M\sin{i}$ would correspond to a true dynamical mass of $34.19\pm21.72$, in full agreement with the NSS true mass estimate.

        \subsection{HIP\,60865}  \label{subsec:gfu16}
            HIP\,60865 is a K7 star at a distance of 72\pc. No long-period modulation is evident from the available ASAS and TESS (Sectors 22 and 49) light curves. The \gdr{3} NSS solution for this star suggests the existence of a companion having an orbital period of $500.69\pm4.88$\days, an eccentricity of $0.25\pm0.06$ and, by assuming a host star mass of 0.67\Msun, lower $M_{\rm NSS}\sim$46.97\Mjup.
            Although five archival SOPHIE measurements are available, their $\sim$0.1\kmps uncertainty makes them unsuitable for RV orbital fitting in the presence of our high-precision HARPS-N measurements and are therefore not included in our analysis.
            \par The GLS periodogram computed for our HARPS-N datapoints, shown in the top row of Fig.~\ref{fig:GFU16-timeseries}, exhibits a principal peak of $\sim$620\days with a highly significant FAP of 1.1$\cdot10^{-14}$\%. While the majority of the activity indices (third-to-eighth row of Fig.~\ref{fig:GFU16-timeseries}) do not show either significant peaks or correlations with the RVs, an exception is represented by the H$\alpha$ index, having $\rho=-0.91$ with a zero p-value and for which we find a significant (FAP=1.6$\cdot10^{-4}$\%) peak at the same period of the aforementioned RV signal. Although this coinciding periodicity could in principle indicate an activity origin for the proposed signal, we must note that the very large variation observed in the RV time series ($\sim$2835\mps peak-to-valley) is incompatible with any expected activity-induced Doppler signal for such an old star. Moreover, the compatibility between the RV periodogram principal peak and the $\sim$500\days orbital period proposed by the \gdr{3} NSS solution, which is by its nature mostly insensitive to signals induced by stellar activity, suggests a companion origin for the observed RV variation. As such, HIP\,60865 may have an activity cycle coinciding with the orbital period of its massive companion, a situation similar to that of the Solar System, in which the orbital period of Jupiter coincides with the $\sim$11\yr solar sunspot cycle (which itself produces a RV variation of $4.98\pm1.44$\mps, as detailed in \citealt{lanza2016}).
            \par We find the data to be best described by a Keplerian solution having an orbital period of $528.08\pm0.39$\days, semi-amplitude of $1418.62_{-1.76}^{+1.81}$\mps and eccentricity of $0.261\pm0.002$. By using the $0.683\pm0.026$\Msun value for host star mass obtained by our SED fit (see Sect.~\ref{sec:star-parameters}) we are able to derive a minimum mass of $43.93\pm1.09$\Mjup and a semi-major axis value of $1.15\pm0.01$\au for the companion. We find no significant signal in the residual RVs (second row of Fig.~\ref{fig:GFU16-timeseries}), with the principal peak at $\sim$1\days having a 0.3\% FAP, as well as no correlation with the activity indices considered. Once again, our RV solution is found to be in strong agreement with the astrometric candidate proposed in \gdr{3}, with the notable exception of a larger RV-only orbital period within 5.6$\sigma$ of the NSS candidate value.
            \par  This star is also included in the HGCA, and we therefore perform a combined RV ad PMa fit with \texttt{orvara} as done for \ref{subsec:gfu25}, finding for the detected companion a semi-major axis of $1.15\pm0.2$\au, eccentricity of $0.260\pm0.002$, true dynamical mass of ${46.1}_{-2.2}^{+3.8}$\Mjup and orbital inclination of ${76}_{-13}^{+18}$ deg agreeing with $i_{\rm NSS}$ within 0.4$\sigma$.
            
        \subsection{PM\,J13580+3141}  \label{subsec:gfu19}
            PM\,J13580+3141 is a K7 star having an estimated distance of 73\pc from the Sun. No long-term stellar modulation is evident from the ASAS and TESS (Sectors 23 and 50) data, but we can infer from ASAS a rotation period of $13.2\pm0.3$\days, while from TESS we obtain $7.0_{-0.5}^{+0.7}$\days, nearly half of the ASAS value, and a similar short-period variability (6.6\days) is present in the Gaia photometry. The NSS solution is compatible with a companion with lower $M_{\rm NSS}\sim$11.66\Mjup (assuming a host star mass of 0.64\Msun) on a $564.02\pm11.19$\days orbit with an eccentricity of $0.51\pm0.19$.
            \par This candidate companion was first validated as Gaia-4\,b by \cite{stefansson2025} by jointly fitting the Gaia astrometric orbit and a total of 23 RV datapoints collected with the HPF \citep{mahadevan2014}, NEID \citep{schwab2016} and FIES \citep{telting2014} spectrographs. These RV measurements allowed for the validation of the astrometric candidate, which was found to be a giant planet having a true mass of $11.80_{-0.66}^{+0.73}$\Mjup on a $571.3_{-1.3}^{+1.4}$\days orbit with an eccentricity of $0.338_{-0.023}^{+0.026}$, in perfect agreement with the astrometric candidate orbit.
            \par Our HARPS-N time series exhibit a clear Keplerian-like variation, with its periodogram (see Fig.~\ref{fig:GFU19-timeseries}) featuring a highly significant (FAP=1.3$\cdot10^{-9}$\%) long-periodicity region starting at $\sim$500\days, clearly compatible with the long-period planet Gaia-4\,b. Our HARPS-N time series covers a significant portion of the orbit of Gaia-4\,b ($\sim$96\%) and is characterized by a significantly lower median uncertainty than the $\sim$12\mps median error of the RV time series analysed in \cite{stefansson2025}, allowing for an updated characterization of the planetary companion orbit.
            \par Combining our HARPS-N time series with the literature measurements we find for Gaia-4\,b an updated Keplerian solution (shown in the fifth panel of Fig.~\ref{fig:rvfit}) with orbital period $574.46_{-2.83}^{+3.06}$\days, semi-amplitude $309.97_{-2.47}^{+2.70}$\mps and eccentricity $0.336_{-0.014}^{+0.013}$. While no significant residual periodogram peak is present, the post-fit residual timeseries feature some disagreement between the RV model and the first few HPF and FIES datapoints, resulting in significant residual scatter in the first descending portion of the RV model. Such scatter could, in principle, be produced by an outer massive companion imposing a long-term trend in the RV data. As such, we perform an additional orbital fit in which we add a quadratic trend to the Keplerian model, resulting in a solution that is marginally statistically preferred (see sixth panel of Fig.~\ref{fig:rvfit}), with a difference in the Bayesian Information Criterion of 9.11. As such, we adopt this solution, in which we find an orbital period of $588.58_{-2.50}^{+2.15}$\days, semi-amplitude $347.42_{-2.75}^{+2.15}$\mps and eccentricity $0.318_{-0.012}^{+0.011}$, with quadratic trend coefficients of $-0.21\pm0.06\,\rm{m\,s}^{-1}\,\rm{d}^{-1}$ and $0.0006\pm0.0001\,\rm{m\,s}^{-1}\,\rm{d}^{-2}$. From the host mass of $0.652^{+0.027}_{-0.025}$\Msun obtained from the SED fit, we derive a minimum planetary mass of $10.30\pm0.45$\Mjup and semimajor axis $1.20\pm0.02$\au. The post-fit residuals have a w.r.m.s. of $\sim$7\mps and no significant periodogram peak.
            \par While we note that the orbital period obtained from our RV-only solution is slightly less constrained than the joint RV and Gaia solution presented in \cite{stefansson2025}, it is instead better constrained than the RV-only value of the cited work ($513_{-16}^{+21}$\days), as a result of both the extended observational timespan and the higher-precision HARPS-N measurement (having average uncertainty of 2.95\mps compared to the 16, 8.3 and 12.4\mps of the HPF, NEID and FIES timeseries respectively). The robust characterization of the quadratic trend and its source requires additional observation, and will therefore be subject of future analysis.

        \subsection{TYC\,3456-251-1} \label{subsec:gfu17}
            TYC\,3456-251-1 is a G9 star located at a distance of 151\pc. The available ASAS and TESS (Sectors 22 and 49) light curves do not exhibit long-term modulation. For this star, the \gdr{3} NSS catalog is compatible with the presence of a massive companion with an orbital period of $593.26\pm7.89$\days and eccentricity $0.14\pm0.07$, which, by assuming a host mass of 0.87\Msun, implies lower $M_{\rm NSS}\sim$52.3\Mjup.
            \par As shown in Fig.~\ref{fig:GFU17-timeseries}, the GLS periodogram for our HARPS-N RV time series is characterized by a long-period ($>$400\days) signal with a FAP of 1.4$\cdot10^{-11}$\%, consistent with the proposed astrometric candidate. The only activity index exhibiting strong correlation with the RV time series is Na\,I ($\rho=-0.68$ and zero p-value), for which we find a peak periodicity of $\sim$488\days with a FAP of 1$\cdot10^{-6}$\%. As discussed in Sections \ref{subsec:gfu25} and \ref{subsec:gfu16} for stars HIP\,105707 and HIP\,60865, we refute an activity-only origin for the aforementioned long-period signal by virtue of its large variation of the RV time series ($\sim$3200\mps) and the fact that the periodicity of the RV signal is consistent with the proposed astrometric orbital solution.
            \par Performing an orbital fit of the data, we find a best-fit Keplerian solution with an orbital period of $592.66_{-4.31}^{+4.92}$\days, semi-amplitude $1662.35_{-15.22}^{+17.60}$\mps and a moderate eccentricity of $0.387\pm0.003$. This best-fit solution, using the host star mass of $0.854^{+0.050}_{-0.041}$\Msun obtained by our SED fit, would correspond to a minimum companion mass of $59.52_{-2.38}^{+2.46}$\Mjup and a semimajor axis of $1.34\pm0.03$\au. We find our solution to be in high agreement with the \gdr{3} NSS candidate solution, with the notable exceptions of significantly larger values of eccentricity and argument of periastron from our solution, being respectively within 3.5$\sigma$ and 6.2$\sigma$ of the NSS solution. While the post-fit residual time series exhibit a relatively high scatter, with a w.r.m.s. of 11\mps, no significant peak is present in the current data, with the principal residual peak being located at $\sim$8.6\days and having a FAP of 0.8\%.
        
        \subsection{TYC\,4998-437-1} \label{subsec:gfu20}
            TYC\,4998-437-1 is a K1 star at a distance of 132\pc from the Sun. No long-period modulation is clear from the ASAS and TESS (Sectors 51 and 91) light curves. For this star, the \gdr{3} NSS solution suggests a candidate companion on a $526.14\pm23.10$\days orbit with an eccentricity of $0.43\pm0.16$, from which lower $M_{\rm NSS}\sim$33.61\Mjup can be estimated assuming a mass of 0.81\Msun for the host star \citep{gaiabinmass2022}.
            \par Although the HARPS-N RV timeseries (top row of Fig.~\ref{fig:GFU20-timeseries}) shows a very clear variation hinting at a Keplerian-like long-period behaviour, the corresponding GLS periodogram lacks any highly significant peak, with its 495\days principal peak having a FAP of 0.20\% being nevertheless compatible with the astrometric candidate orbit. We find moderate correlation between RVs and most of the activity indices, with the significant exception of the strong correlation found with H$\alpha$ ($\rho=0.83$ with a zero p-value). However, we once again stress that the amplitude of the observed RV variation ($\sim$1433\mps peak-to-valley) is too large to be induced by activity alone, with the astrometric candidate additionally motivating the search for a companion causing the RV variation.
            \par The orbital fit we perform on the data results in a best-fit Keplerian with a $558.67_{-2.47}^{+2.80}$\days period, $738.92_{-1.87}^{+1.75}$\mps semi-amplitude and $0.557\pm0.002$ eccentricity. From this, using our SED-derived host star mass of $0.856^{+0.043}_{-0.035}$\Msun, we find a companion minimum mass value of $22.80_{-0.77}^{+0.75}$\Mjup and semi-major axis of $1.27\pm0.02$\au. Once again, we note the high agreement between our RV solution and the astrometric candidate orbit, and no significant peak is present in the residual timeseries.
            
        \subsection{StKM\,1-1494} \label{subsec:gfu22}
            StKM\,1-1494 is a M0 star at a distance of 27\pc. No TESS data are available; however, we infer a possible rotational period of $6\pm3$\days from the ASAS light curve and a $\sim$12\days periodicity from Gaia photometry, which may be aliases of each other. The \gdr{3} solution has an orbital period of $829.40\pm1.76$\days and an eccentricity of $0.328\pm0.006$, corresponding to lower $M_{\rm NSS}\sim$64.2\Mjup assuming a host mass of 0.58\Msun.
            \par This candidate companion has recently been confirmed through direct imaging observations performed with the GRAVITY \citep{abuter2017} near-infrared interferometer as detailed in \cite{winterhalder2024}. GRAVITY allowed for the direct imaging of the candidate companion, which combined with the Gaia orbital solution allowed its validation and characterization as having an orbital period of $828.2\pm1.6$\days, eccentricity of $0.331\pm0.006$ and a mass of $67.5\pm1.2$\Mjup. We additionally note that this star additionally has a SB1 solution in DR3, although the short period of 7.9 days and large amplitudes in excess of 1.8 km/s suggest incorrect identification of the correct period value due to, e.g., aliasing effects.
            \par Our HARPS-N observations cover $\sim$62\% of the candidate orbital period, therefore allowing only the detection of a large trend in the data with only recent hints of curvature, with a total RV variation of 3584\mps and producing a significant GLS periodogram long-period ($>$800\days) signal with a 3$\cdot10^{-27}$\% FAP. Although a robust solution would clearly require a larger orbital coverage, the most recent data manage to sample the minimum of the RV oscillation, allowing us to perform an RV fit to provide a first assessment for the spectroscopic orbit of the detected brown dwarf. We find a best-fit solution having an orbital period of $789.26_{-11.25}^{+10.08}$\days, semi-amplitude $1984.95_{-18.19}^{+11.16}$\mps and a $0.335\pm0.007$ eccentricity. We note that our solution is characterized by a significantly shorter period than the NSS candidate one (within 4.15$\sigma$). This disagreement is likely a result of the incomplete coverage of the orbital period provided by our HARPS-N data, highlighting again how further observations can improve the orbital characterization of this companion. From the host star mass of $0.584\pm0.027$\Msun we infer a minimum companion mass of $63.35_{-2.03}^{+1.93}$\Mjup and semimajor axis of $1.44\pm0.02$\au. The residual time series features a principal peak of $\sim$15\days with a $0.01$\% FAP coinciding with significant peaks in the FWHM, Ca II, H$\alpha$, suggesting an activity origin for the residual RV scatter.

    \section{Discussions and conclusions}  \label{sec:conclusions}
        In this work we have presented the first results of an ongoing RV survey conducted with HARPS-N on 35 stars hosting cold Jupiters or low-mass brown dwarfs. Specifically, we have focused on the 14 stars in the sample characterized by \gdr{3} non-single-star solutions compatible with the presence of such massive companions, in order to identify astrophysical false positives in the NSS solutions and provide RV confirmation and characterization of the remaining validated candidates.
        \par We have identified 6 NSS solutions in the sample as astrophysical false positives originating from unresolved SB2 mimicking substellar astrometric motion, from which we can estimate a $43_{-11}^{+13}$\% binary contamination fraction in our sample. Notably, this value is higher than preliminary contamination estimations reported in the literature, such as the $\sim$10\% derived in \cite{holl2023}, but is instead in agreement with more recent and higher value of $\sim$32\% \citep[see e.g.][]{stefansson2025}. This result highlights the critical role of RV follow-up observations in validating astrometric candidates.
        We additionally note that these false positives tend to be the stars at largest distances in our sample. Indeed, five out of the six targets with parallaxes lower than 7.5\mas in the sample are found to be binaries. Moreover, five out of six false positives being found around stars having masses above 0.9\Msun, in line with the frequency of binaries growing with the primary mass \citep[see e.g.][]{moe2017}.
        \par The HARPS-N data collected in our survey allowed us to validate and confirm the remaining 8 astrometric substellar candidates in our sample, finding RV best-fit solutions compatible with giant and brown dwarf companions having minimum masses ranging from 8\Mjup to 62\Mjup and semimajor axes from 0.75\au to 1.42\au. While we additionally note that two companions have recently been confirmed by other observational efforts, namely RV for PM~J13580+3141 \citep{stefansson2025} and imaging data for StKM 1-149 \citep{winterhalder2024}, our high-precision HARPS-N measurements allow for updated orbital solutions. In general, our solutions are in good agreement with the NSS astrometric solutions, and by adopting the NSS orbital inclination to derive true mass estimates from our $M\sin{i}$ we obtain values higher than or compatible with the NSS minimum companion mass. 
        \begin{figure}
            \centering
            \includegraphics[width=\linewidth]{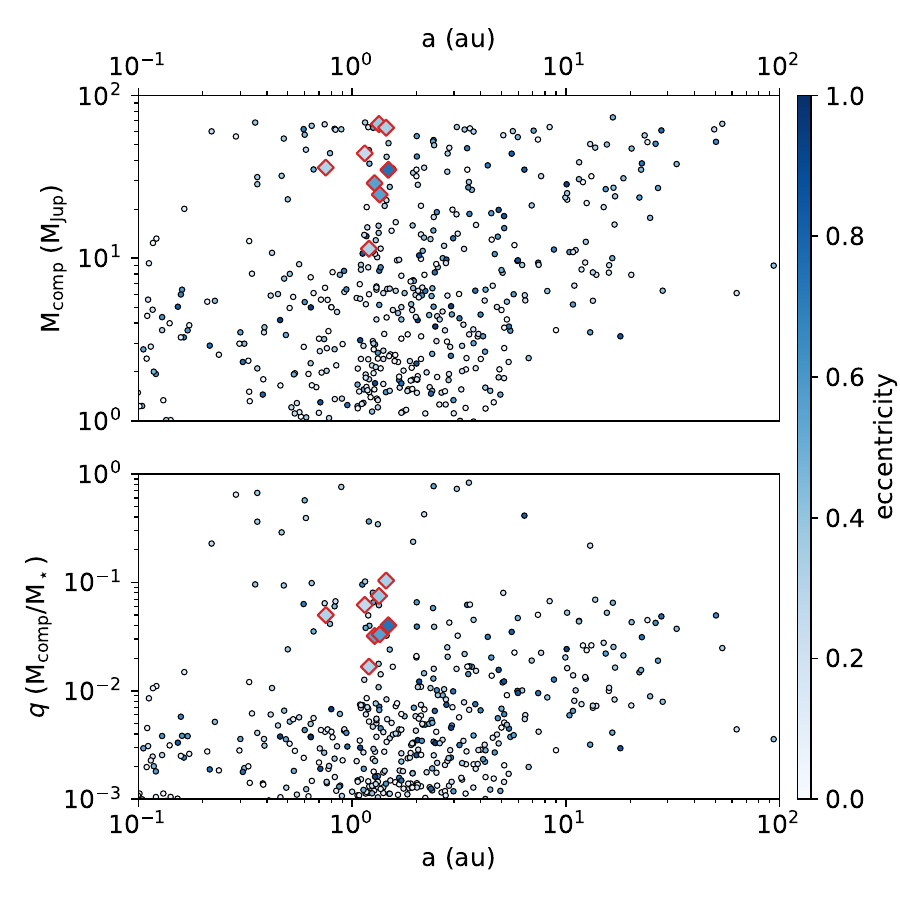}
            \caption{Location of the eight companion discussed in this work (red-bordered diamonds) in the known exoplanet and brown dwarf (black-bordered circles) single-companion population, color-coded according to orbital eccentricity. \textit{Top panel}: companion mass vs orbital separation. \textit{Bottom panel}: companion-to-host mass ratio $q$ vs orbital separation.}
            \label{fig:exocatalog}
        \end{figure}
        \par Finally, in Fig.~\ref{fig:exocatalog} we show the location of the eight companions analysed in this work, represented with their true mass estimates $M_{b,i_{\rm NSS}}$, in the known single-companion brown dwarf and exoplanetary population with available true mass estimates, from which we can note that the more massive of these companions are found to be populating the brown dwarf desert \citep[$<$5\au, see][]{marcy2000,sahlmann2011}.
        \par As an additional testament to the continued impact of Gaia astrometric measurements on other techniques, we note how some of the astrometric candidate solutions discussed in this work have orbital periods compatible with activity cycles as detected by the spectroscopical analysis performed on our HARPS-N data, specifically for stars HIP~60865, HIP~105707, TYC~3456-251-1, and TYC~4998-437-1. As the astrometric method is by its nature mostly insensitive to signals induced by stellar activity, the astrometric identification of a candidate orbital period can be a key asset in refuting a stellar activity origin for observed RV signals. Although the Keplerian nature of the observed RV signals presented in this work is clear from their very large RV amplitudes alone, we can foresee a significant role in the near future for Gaia astrometric solutions in solving more dubious cases such as those with smaller RV amplitudes, which by itself could easily remain ambiguous, as further RV follow-up of Gaia candidates is conducted.
        \par While the present work has focused solely on the validation and confirmation of astrometric candidates in our 35-star HARPS-N sample, a final key objective of our survey, namely the search for low-mass inner companions in the presence of outer massive companions with astrometric-derived true masses, is ongoing and will be the subject of forthcoming papers. As the search for inner companions continues within our stellar sample, we will be able to provide an updated assessment of ordered system demographics based on outer companion true masses as provided by astrometry, highlighting once more the growing role of multi-technique observations in the exoplanetary landscape.

    \section{Data availability}
        Full Table \ref{table:rvdata} is available in electronic form at the CDS via anonymous ftp to \url{cdsarc.u-strasbg.fr} (\url{130.79.128.5}) or via \url{http://cdsweb.u-strasbg.fr/cgi-bin/qcat?J/A+A/}.
                
    \begin{acknowledgements}
        The authors wish to thank the referee, Dr. C. Babusiaux, for the useful comments which significantly improved the manuscript quality.
        This work is based on observations made with the Italian \textit{Telescopio Nazionale Galileo} (TNG), operated on the island of La Palma by the INAF - \textit{Fundaci\'on Galileo Galilei} at the \textit{Roque de Los Muchachos} Observatory of the \textit{Instituto de Astrof\'isica de Canarias} (IAC).
        We acknowledge the Italian center for Astronomical Archives (IA2, \url{https://www.ia2.inaf.it}), part of the Italian National Institute for Astrophysics (INAF), for providing technical assistance, services and supporting activities of the GAPS collaboration.
        This work has made use of data from the European Space Agency (ESA) mission {\it Gaia} (\url{https://www.cosmos.esa.int/gaia}), processed by the {\it Gaia} Data Processing and Analysis Consortium (DPAC, \url{https://www.cosmos.esa.int/web/gaia/dpac/consortium}). Funding for the DPAC has been provided by national institutions, in particular, the institutions participating in the {\it Gaia} Multilateral Agreement. 
        This research has made extensive use of the NASA-ADS, SIMBAD and Vizier databases, operated at CDS, Strasbourg, France.
        M.P. acknowledges support from the European Union – NextGenerationEU (PRIN MUR 2022 20229R43BH), the ``Programma di Ricerca Fondamentale INAF 2023'', and from ASI-INAF agreement no. 2025-10-HH.0 ``Partecipazione Italiana al Gaia DPAC - Supporto alle attivit\`a di responsabilit\`a del team scientifico''.
        J.M. acknowledges support from the Italian Ministero dell’Università e della Ricerca and the European Union – Next Generation EU through project PRIN 2022 PM4JLH “Know your little neighbours: characterising low-mass stars and planets in the Solar neighbourhood”.
        L.N. acknowledges financial contribution from the INAF Large Grant 2023 ``EXODEMO''. 
        L.M. acknowledges the financial contribution from the PRIN MUR 2022 project 2022J4H55R.
    \end{acknowledgements}
    
    \bibliographystyle{aa}
    \bibliography{ref}
    
    \appendix
    \section{Stellar and orbital parameters}
        \begin{sidewaystable*}
            \caption{Stellar parameters for Gaia non-single-star candidates in the sample identified as astrophysical false positives in this work.}       \label{table:star-binary}
            \centering
            \begin{tabular}{l c c c c c c}
                \hline\hline
                    &TYC\,4338-562-1            &HD\,275921                 &TYC\,4730-512-1            &TYC\,4106-138-1            &PM\,J08079+2651                &TYC\,1546-519-1\\
                \hline
                    \gdr{3} ID                              &547770821740800384         &230138989266819456         &3205578257602278784        &288196433028185728         &683525873153063680             &4553792062596291072\\[3pt]
                    $\alpha$ (J2000)\tablefootmark{a}       &3\fh16\fm32.83\fs          &3\fh56\fm18.52\fs          &4\fh27\fm7.03\fs           &5\fh54\fm58.09\fs          &8\fh07\fm55.53\fs              &17\fh29\fm39.40\fs\\[3pt]
                    $\delta$ (J2000)\tablefootmark{a}       &+74\fdg13\fmin51.14\fsec   &+40\fdg51\fmin44.04\fsec   &-2\fdg55\fmin50.68\fsec    &+65\fdg47\fmin1.43\fsec    &+26\fdg51\fmin31.46\fsec       &+18\fdg47\fmin38.14\fsec\\[3pt]
                    \parallax (\mas)\tablefootmark{a}       &$4.819\pm0.015$          &$6.082\pm0.027$          &$4.591\pm0.023$          &$7.555\pm0.029$          &$21.483\pm0.032$             &$5.150\pm0.0160$\\[3pt]
                    $\mu_\alpha$ (\masyr)\tablefootmark{a}  &$-5.282\pm0.015$           &$-22.885\pm0.029$          &$52.469\pm0.022$         	&$-28.600\pm0.019$          &$39.641\pm0.035$               &$24.153\pm0.077$\\[3pt]
                    $\mu_\delta$ (\masyr)\tablefootmark{a}  &$-10.059\pm0.018$          &$-10.608\pm0.031$          &$49.380\pm0.018$           &$-36.909\pm0.019$          &$68.794\pm0.032$               &$26.456\pm0.056$\\[3pt]
                    RUWE \tablefootmark{a}                  &1.50                       &1.54                       &1.42                       &2.10                       &2.63                           &1.47\\[3pt]
                    $RV_{\rm sys}$ (\kmps)\tablefootmark{a} &$4.12\pm0.49$              &$-13.72\pm1.27$            &$101.41\pm0.84$            &$3.49\pm0.24$              &$8.65\pm0.01$                  &$-27.38\pm0.43$\\[3pt]
                    $P_{\rm NSS}$ (\days)\tablefootmark{a}  &$512.48\pm7.78$            &$739.78\pm31.77$           &$598.12\pm19.31$           &$438.23\pm7.05$            &$113.62\pm0.40$                &$988\pm119$\\[3pt]
                    $B$ (mag)\tablefootmark{b}              &$10.40\pm0.03$             &$10.77\pm0.05$             &$11.33\pm0.09$             &$12.09\pm0.12$             &$13.74\pm0.05$                 &$12.06\pm0.12$\\[3pt]
                    $V$ (mag)\tablefootmark{b}              &$9.87\pm0.03$              &$10.08\pm0.04$             &$10.60\pm0.07$             &$11.16\pm0.06$             &$12.24\pm0.02$                 &$11.18\pm0.07$\\[3pt]
                    $J$ (mag)\tablefootmark{c}              &$8.886\pm0.030$            &$8.934\pm0.056$            &$9.543\pm0.027$            &$9.601\pm0.022$            &$9.166\pm0.021$                &$9.863\pm0.020$\\[3pt]
                    $H$ (mag)\tablefootmark{c}              &$8.728\pm0.030$            &$8.668\pm0.023$            &$9.2\pm0.026$              &$9.139\pm0.029$            &$8.545\pm0.023$                &$9.522\pm0.016$\\[3pt]
                    $K$ (mag)\tablefootmark{c}              &$8.643\pm0.022$            &$8.593\pm0.019$            &$9.133\pm0.023$            &$9.050\pm0.024$            &$8.354\pm0.017$                &$9.462\pm0.018$\\[3pt]
                    WISE1 (mag)\tablefootmark{d}            &$8.621\pm0.023$            &$8.545\pm0.023$            &$9.064\pm0.022$            &$9.015\pm0.023$            &$8.240\pm0.022$                &$9.419\pm0.024$\\[3pt]
                    WISE2 (mag)\tablefootmark{d}            &$8.654\pm0.019$            &$8.571\pm0.021$            &$9.113\pm0.019$            &$9.068\pm0.020$            &$8.165\pm0.020$                &$9.465\pm0.020$\\[3pt]
                    WISE3 (mag)\tablefootmark{d}            &$8.615\pm0.023$            &$8.535\pm0.026$            &$9.066\pm0.029$            &$9.027\pm0.030$            &$8.091\pm0.024$                &$9.400\pm0.034$\\[3pt]
                    WISE4 (mag)\tablefootmark{d}            &$8.602\pm0.412$            &$8.010\pm0.211$            &$8.318\pm0.010$            &$8.925\pm0.467$            &$8.046\pm0.251$                &$8.995\pm0.448$\\[3pt]
                    \gmag (mag)\tablefootmark{a}            &$9.845\pm0.002$            &$9.958\pm0.003$            &$10.595\pm0.003$           &$10.951\pm0.003$           &$11.449\pm0.002$               &$11.019\pm0.003$\\[3pt]
                    \gbp (mag)\tablefootmark{a}             &$10.110\pm0.003$           &$10.280\pm0.003$           &$10.941\pm0.003$           &$11.400\pm0.003$           &$12.468\pm0.003$               &$11.381\pm0.003$\\[3pt]
                    \grp (mag)\tablefootmark{a}             &$9.416\pm0.004$            &$9.466\pm0.004$            &$10.081\pm0.004$           &$10.335\pm0.004$           &$10.452\pm0.004$               &$10.486\pm0.004$\\[3pt]
                    Spectral Type                           &F8                         &G5                         &G6                         &K0                         &K7                             &K0\\[3pt]
                    $M_\star$ (\Msun)\tablefootmark{e}      &$1.46^{+0.13}_{-0.14}$     &$1.18^{+0.13}_{-0.15}$     &$1.20^{+0.17}_{-0.18}$     &$0.960^{+0.072}_{-0.064}$  &$0.641\pm0.027$                &$1.030^{+0.096}_{-0.097}$\\[3pt]
                    $R_\star$ (\Rsun)\tablefootmark{e}      &$1.667^{+0.085}_{-0.077}$  &$1.374^{+0.065}_{-0.062}$  &$1.64^{+0.13}_{-0.12}$     &$0.987^{+0.043}_{-0.038}$  &$0.622^{+0.021}_{-0.020}$      &$1.127^{+0.050}_{-0.047}$\\[3pt]
                    $\rho_\star$ (\gcm)\tablefootmark{e}    &$0.442^{+0.087}_{-0.080}$  &$0.64^{+0.13}_{-0.12}$     &$0.384^{+0.11}_{-0.093}$   &$1.41^{+0.20}_{-0.18}$     &$3.76^{+0.33}_{-0.31}$         &$1.01^{+0.17}_{-0.15}$\\[3pt]
                    $L_\star$ (\Lsun)\tablefootmark{e}      &$5.4^{+1.4}_{-1.0}$        &$2.62^{+0.56}_{-0.40}$     &$3.28^{+0.88}_{-0.66}$     &$0.873^{+0.11}_{-0.100}$   &$0.0810^{+0.0069}_{-0.0059}$   &$1.41^{+0.22}_{-0.18}$\\[3pt]
                    Age (Gyr)\tablefootmark{e}              &$1.41^{+1.3}_{-0.88}$      &$3.5^{+3.9}_{-2.4}$        &$4.4^{+4.1}_{-2.4}$        &$6.4^{+4.5}_{-4.1}$        &$7.4^{+4.4}_{-4.8}$            &$5.4^{+4.6}_{-3.5}$\\[3pt]
                    \teff (K)\tablefootmark{e}              &$6800^{+440}_{-400}$       &$6270^{+340}_{-290}$       &$6090^{+350}_{-340}$       &$5610^{+150}_{-140}$       &$3907^{+58}_{-56}$             &$5930^{+220}_{-210}$\\[3pt]
                    \logg (cgs)\tablefootmark{e}            &$4.157^{+0.061}_{-0.069}$  &$4.234^{+0.064}_{-0.080}$  &$4.090^{+0.083}_{-0.098}$  &$4.431\pm0.045$            &$4.657^{+0.024}_{-0.025}$      &$4.347^{+0.053}_{-0.058}$\\[3pt]
                    \feh\tablefootmark{e}                   &$-0.05^{+0.25}_{-0.30}$    &$-0.09^{+0.20}_{-0.25}$    &$-0.10^{+0.27}_{-0.28}$    &$0.07^{+0.18}_{-0.16}$     &$0.31^{+0.12}_{-0.15}$         &$-0.02^{+0.21}_{-0.23}$\\[3pt]
                \hline
            \end{tabular}
            \tablefoot{
                \tablefoottext{a}{retrieved from Gaia Data Release 3 \citep{vallenari2023}}
                \tablefoottext{b}{retrieved from \cite{chambers2016}}
                \tablefoottext{c}{retrieved from \cite{cutri2003}}
                \tablefoottext{d}{retrieved from \cite{cutri2014}}
                \tablefoottext{e}{obtained from the SED fitting discussed in Sect. \ref{sec:star-parameters}}
                }
        \end{sidewaystable*}
        \begin{sidewaystable*}
            \caption{Stellar parameters for Gaia non-single-star candidates in the sample validated and confirmed in this work.}       \label{table:star-validated}
            \centering
            \fontsize{7.2}{4}\selectfont
            \begin{tabular}{l c c c c c c c c}
                \hline\hline
                    &2MASS\,J20570943+1406590   &HIP\,105707                &TYC\,198-493-1               &HIP\,60865                &PM\,J13580+3141            &TYC\,3456-251-1           &TYC\,4998-437-1           &StKM\,1-1494\\
                \hline
                    \gdr{3} ID                                &1761619456801656064        &1853704144046204032        &3087990814774129920          &1518957932040718464       &1457486023639239296        &1543883104727682176       &6330529666839726592       &4169462136799175296\\[3pt]
                    $\alpha$ (J2000)\tablefootmark{a}         &20\fh57\fm9.44\fs          &21\fh24\fm36.60\fs         &8\fh2\fm5.21\fs              &12\fh28\fm30.86\fs        &13\fh58\fm01.62\fs         &12\fh33\fm40.86\fs        &14\fh46\fm31.25\fs        &17\fh34\fm46.72\fs\\[3pt]
                    $\delta$ (J2000)\tablefootmark{a}         &+14\fdg06\fmin59.03\fsec   &+32\fdg07\fmin22.76\fsec   &+2\fdg36\fmin44.63\fsec      &+35\fdg16\fmin36.79\fsec  &+31\fdg41\fmin43.48\fsec   &+48\fdg38\fmin39.28\fsec  &-7\fdg24\fmin34.41\fsec   &-06\fdg54\fmin10.72\fsec\\[3pt]
                    \parallax (\mas)\tablefootmark{a}         &$9.080\pm0.019$            &$19.871\pm0.023$           &$13.297\pm0.035$             &$13.395\pm0.026$          &$13.643\pm0.025$           &$6.485\pm0.024$           &$7.673\pm0.041$           &$35.793\pm0.024$\\[3pt]
                    $\mu_\alpha$ (\masyr)\tablefootmark{a}    &$19.517\pm0.022$           &$-156.591\pm0.034$         &$-35.016\pm0.041$            &$-278.109\pm0.023$        &$-75.528\pm0.023$          &$-36.293\pm0.019$         &$-94.926\pm0.036$         &$-8.747\pm0.034$\\[3pt]
                    $\mu_\delta$ (\masyr)\tablefootmark{a}    &$-47.591\pm0.017$          &$-116.689\pm0.023$         &$-21.455\pm0.021$            &$2.888\pm0.024$           &$18.101\pm0.017$           &$-12.127\pm0.022$         &$34.270\pm0.031$          &$-74.531\pm0.025$\\[3pt]
                    $U$ (\kmps)\tablefootmark{b}              &$-11.055\pm0.199$          &$-43.670\pm0.075$          &$13.371\pm0.291$             &$83.461\pm0.176$          &$26.053\pm0.043$           &$12.884\pm0.171$          &$79.715\pm1.092$          &$-1.193\pm0.454$\\[3pt]
                    $V$ (\kmps)\tablefootmark{b}              &$-9.049\pm0.364$           &$-17.177\pm0.297$          &$-12.025\pm0.234$            &$-49.466\pm0.100$         &$-15.547\pm0.023$          &$-27.470\pm0.196$         &$-14.497\pm0.294$         &$-9.191\pm0.145$\\[3pt]
                    $W$ (\kmps)\tablefootmark{b}              &$-23.381\pm0.156$          &$9.698\pm0.0.070$          &$-9.813\pm0.120$             &$-20.128\pm0.405$         &$-10.428\pm0.014$          &$-20.694\pm0.594$         &$-0.159\pm1.139$          &$-4.467\pm0.115$\\[3pt]
                    RUWE \tablefootmark{a}                    &$1.97$                     &$2.819$                    &$3.26$                       &$3.84$                    &$1.16$                     &$2.68$                    &$1.63$                    &$3.15$\\[3pt]
                    $RV_{\rm sys}$ (\kmps)\tablefootmark{a}   &$5.4402\pm0.4397$          &$-11.21\pm0.31$            &$14.33\pm0.39$               &$-11.3957\pm0.4102$       &$-17.348\pm0.008$          &$-23.80\pm0.64$           &$-51.74\pm1.57$           &$-2.66\pm0.49$\\[3pt]
                    $B$ (mag)\tablefootmark{c}                &$13.23\pm0.03$             &$12.66\pm0.17$             &$11.33\pm0.11$               &$14.16\pm0.49$            &$13.82\pm0.02$             &$12.56\pm0.16$            &$12.15\pm0.26$            &$12.80\pm0.28$\\[3pt]
                    $V$ (mag)\tablefootmark{c}                &$12.26\pm0.05$             &$11.20\pm0.06$             &$10.58\pm0.08$               &$12.256\pm0.20$           &$12.52\pm0.02$             &$11.76\pm0.11$            &$11.31\pm0.15$            &$11.24\pm0.01$\\[3pt]
                    $J$ (mag)\tablefootmark{d}                &$10.27\pm0.02$             &$8.75\pm0.02$              &$8.91\pm0.02$                &$9.75\pm0.02$             &$9.94\pm0.02$              &$10.23\pm0.02$            &$10.03\pm0.02$            &$8.30\pm0.02$\\[3pt]
                    $H$ (mag)\tablefootmark{d}                &$9.65\pm0.03$              &$8.11\pm0.02$              &$8.49\pm0.03$                &$9.12\pm0.04$             &$9.30\pm0.01$              &$9.86\pm0.02$             &$9.58\pm0.02$             &$7.71\pm0.03$\\[3pt]
                    $K$ (mag)\tablefootmark{d}                &$9.59\pm0.01$              &$7.98\pm0.02$              &$8.41\pm0.02$                &$9.01\pm0.02$             &$9.13\pm0.02$              &$9.79\pm0.02$             &$9.52\pm0.02$             &$7.45\pm0.027$\\[3pt]
                    WISE1 (mag)\tablefootmark{e}              &$9.52\pm0.02$              &$7.91\pm0.02$              &$8.37\pm0.02$                &$8.92\pm0.02$             &$9.05\pm0.02$              &$9.76\pm0.02$             &$9.44\pm0.02$             &$7.34\pm0.03$\\[3pt]
                    WISE2 (mag)\tablefootmark{e}              &$9.59\pm0.02$              &$7.99\pm0.02$              &$8.41\pm0.02$                &$8.97\pm0.02$             &$9.11\pm0.02$              &$9.80\pm0.02$             &$9.495\pm0.02$            &$7.34\pm0.02$\\[3pt]
                    WISE3 (mag)\tablefootmark{e}              &$9.49\pm0.03$              &$7.89\pm0.02$              &$8.36\pm0.02$                &$8.89\pm0.03$             &$8.99\pm0.03$              &$9.75\pm0.04$             &$9.411\pm0.03$            &$7.27\pm0.02$\\[3pt]
                    WISE4 (mag)\tablefootmark{e}              &$8.44\pm0.02$              &$7.64\pm0.13$              &$8.50\pm0.38$                &$8.71\pm0.38$             &$8.86\pm0.35$              &$8.53\pm0.01$             &$8.817\pm0.03$            &$7.16\pm0.10$\\[3pt]
                    \gmag (mag)\tablefootmark{a}              &$11.88\pm0.02$             &$10.568\pm0.003$           &$10.162\pm0.003$             &$11.622\pm0.003$          &$11.909\pm0.003$           &$11.401\pm0.003$          &$11.421\pm0.003$          &$10.535\pm0.003$\\[3pt]
                    \gbp (mag)\tablefootmark{a}               &$12.46\pm0.02$             &$11.298\pm0.003$           &$10.588\pm0.003$             &$12.361\pm0.003$          &$12.723\pm0.003$           &$11.779\pm0.003$          &$11.893\pm0.003$          &$11.520\pm0.003$\\[3pt]
                    \grp (mag)\tablefootmark{a}               &$11.15\pm0.02$             &$9.740\pm0.004$            &$9.571\pm0.003$              &$10.784\pm0.004$          &$11.025\pm0.004$           &$10.854\pm0.004$          &$10.787\pm0.04$           &$9.553\pm0.003$\\[3pt]
                    Spectral Type                             &K4                         &K6/K7                      &K0                           &K7                        &K7                         &G9                        &K1                        &M0\\[3pt]
                    \teff (K)\tablefootmark{f}                &$4590\pm75$                &$4201\pm45$                &$5250\pm90$                  &$4142\pm45$               &$4088\pm45$                &$5400\pm150$              &$5150\pm100$              &$3855\pm45$\\[3pt]
                    \logg (cgs)\tablefootmark{f}              &$4.50\pm0.11$              &$4.66\pm0.02$              &$4.58\pm0.25$                &$4.66\pm0.02$             &$4.66\pm0.02$              &$4.25\pm0.35$             &$4.40\pm0.30$             &$4.68\pm0.02$\\[3pt]
                    \feh\tablefootmark{f}                     &$-0.27\pm0.07$             &$0.01\pm0.11$              &$-0.06\pm0.05$               &$0.02\pm0.11$             &$0.01\pm0.11$              &$-0.18\pm0.05$            &$0.01\pm0.05$             &$-0.01\pm0.11$\\[3pt]
                    v$_{\rm micro}$ (\kmps)\tablefootmark{f}  &$0.80\pm0.05$              &$-$                        &$0.30\pm0.15$                &$-$                       &$-$                        &$0.80\pm0.15$             &$0.35\pm0.10$             &$-$\\[3pt]
                    $M_\star$ (\Msun)\tablefootmark{g}        &$0.689^{+0.031}_{-0.030}$  &$0.707^{+0.028}_{-0.026}$  &$0.817^{+0.037}_{-0.035}$    &$0.683\pm0.026$           &$0.652^{+0.027}_{-0.025}$  &$0.854^{+0.050}_{-0.041}$ &$0.856^{+0.043}_{-0.035}$ &$0.584\pm0.027$\\[3pt]
                    $R_\star$ (\Rsun)\tablefootmark{g}        &$0.687^{+0.033}_{-0.030}$  &$0.710^{+0.018}_{-0.017}$  &$0.774^{+0.026}_{-0.023}$    &$0.679^{+0.018}_{-0.017}$ &$0.641^{+0.020}_{-0.018}$  &$0.879^{+0.069}_{-0.059}$ &$0.871^{+0.050}_{-0.046}$ &$0.571\pm0.022$\\[3pt]
                    $\rho_\star$ (\gcm)\tablefootmark{g}      &$3.00^{+0.38}_{-0.35}$     &$2.78^{+0.20}_{-0.19}$     &$2.49^{+0.24}_{-0.23}$       &$3.08^{+0.23}_{-0.22}$    &$3.48^{+0.28}_{-0.27}$     &$1.78^{+0.38}_{-0.34}$    &$1.83^{+0.30}_{-0.26}$    &$4.43^{+0.44}_{-0.41}$\\[3pt]
                    $L_\star$ (\Lsun)\tablefootmark{g}        &$0.197^{+0.022}_{-0.018}$  &$0.148^{+0.007}_{-0.006}$  &$0.406^{+0.036}_{-0.029}$    &$0.127^{+0.007}_{-0.005}$ &$0.106^{+0.007}_{-0.006}$  &$0.63^{+0.13}_{-0.10}$    &$0.502^{+0.074}_{-0.065}$ &$0.067^{+0.006}_{-0.005}$\\[3pt]
                    Age$_{\rm SED}$ (Gyr)\tablefootmark{g}    &$7.8^{+4.3}_{-6.4}$        &$9.2^{+3.3}_{-5.1}$        &$5.3^{+4.8}_{-3.6}$          &$8.5^{+3.7}_{-4.9}$       &$7.7^{+4.2}_{-4.8}$        &$8.2^{+3.9}_{-4.7}$       &$9.1^{+3.3}_{-4.9}$       &$7.4^{+4.5}_{-4.9}$\\[3pt]
                    Age$_{\rm kin}$ (Gyr)\tablefootmark{b}    &$\sim2.37$                 &$\sim4.92$                 &$\sim0.556$                  &$\sim10.72$               &$\sim0.77$                 &$\sim2.31$                &$\sim5.42$                &$\sim0.60$\\[3pt]
                    \smw\tablefootmark{h}                     &$0.166\pm0.034$            &$0.386\pm0.034$            &$0.195\pm0.035$              &$0.348\pm0.052$           &$0.837\pm0.050$            &$0.150\pm0.012$           &$0.097\pm0.019$           &$1.183\pm0.091$\\[3pt]
                    H$\alpha$\tablefootmark{h}                &$0.165\pm0.008$            &$0.223\pm0.006$            &$0.207\pm0.006$              &$0.225\pm0.013$           &$0.350\pm0.011$            &$0.260\pm0.007$           &$0.447\pm0.012$           &$0.328\pm0.016$\\[3pt]
                    Na I\tablefootmark{h}                     &$0.315\pm0.024$            &$0.299\pm0.010$            &$0.832\pm0.025$              &$0.323\pm0.028$           &$0.425\pm0.018$            &$1.286\pm0.066$           &$1.221\pm0.057$           &$0.312\pm0.011$\\[3pt]
                    He I\tablefootmark{h}                     &$0.557\pm0.005$            &$0.580\pm0.003$            &$0.551\pm0.003$              &$0.574\pm0.003$           &$0.574\pm0.002$            &$0.546\pm0.003$           &$0.552\pm0.006$           &$0.569\pm0.003$\\[3pt]
                \hline
                \end{tabular}
                \tablefoot{
                    \tablefoottext{a}{retrieved from Gaia Data Release 3 \citep{vallenari2023,gaiabinmass2022}}
                    \tablefoottext{b}{obtained from the proper motion analysis discussed in Sect~\ref{sec:star-parameters}}
                    \tablefoottext{c}{retrieved from \cite{chambers2016}}
                    \tablefoottext{d}{retrieved from \cite{cutri2003}}
                    \tablefoottext{e}{retrieved from \cite{cutri2014}}
                    \tablefoottext{f}{obtained from spectroscopical analysis discussed in Sect.~\ref{sec:star-parameters}}
                    \tablefoottext{g}{obtained from the SED fitting discussed in Sect.~\ref{sec:star-parameters}}
                    \tablefoottext{h}{obtained from activity analysis discussed in Sect.~\ref{sec:star-parameters}}
                    }
            \end{sidewaystable*}
            \begin{sidewaystable*}
                \caption{HARPS-N observational statistics, selected \gdr{3} NSS astrometric candidate solution parameters and RV best-fit orbital solutions for the companions detected in this work.}    \label{table:rvfit}
                \centering
                \footnotesize
                \begin{tabular}{l c c c c c c c c}
                    \hline\hline
                        &2MASS\,J20570943+1406590       &HIP\,105707                    &TYC\,198-493-1                 &HIP\,60865                 &PM\,J13580+3141                &TYC\,3456-251-1                &TYC\,4998-437-1                &StKM\,1-1494\\
                    \hline
                        Observation start                   &10/2023                        &10/2023                        &11/2023                        &12/2023                    &01/2024                        &12/2023                        &04/2024                        &04/2024\\[3pt]
                        Observation end                     &12/2024                        &11/2025                        &01/2026                        &04/2025                    &07/2025                        &06/2025                        &07/2025                        &09/2025\\[3pt]
                        $N_{\rm RV}$                        &31                             &59                             &33                             &22                         &17                             &28                             &16                             &35\\[3pt]
                        $t_{\rm span}$ (\days)              &428                            &785                            &796                            &499                        &547                            &538                            &446                            &519\\[3pt]
                        $P_{\rm cov}$ (\%)                  &150                            &114                            &110                            &99                         &96                             &90                             &85                             &62\\[3pt]
                        $\overline{\sigma}_{\rm RV}$ (\mps) &2.38                           &1.36                           &1.45                           &2.28                       &2.95                           &2.46                           &1.97                           &1.93\\[3pt]
                        w.r.m.s. (\mps)                     &544                            &344                            &173                            &999                        &275                            &1187                           &459                            &1318\\[3pt]
                    \hline
                        $P_{\rm NSS}$ (\days)               &$280.76\pm3.16$                &$687.71\pm12.67$               &$722.60\pm8.79$                &$500.69\pm4.88$            &$564.02\pm11.19$               &$592.66\pm7.89$                &$526.14\pm23.10$               &$829.40\pm1.76$\\[3pt]
                        $e_{\rm NSS}$                       &$0.33\pm0.13$                  &$0.51\pm0.03$                  &$0.75\pm0.07$                  &$0.25\pm0.06$              &$0.51\pm0.19$                  &$0.14\pm0.07$                  &$0.43\pm0.16$                  &$0.328\pm0.006$\\[3pt]
                        $i_{\rm NSS}$ (deg)                 &$57.15\pm4.37$                 &$124.06\pm2.08$                &$14.62\pm18.02$                &$82.73\pm1.77$             &$115.43\pm4.60$                &$62.57\pm3.68$                 &$127.69\pm10.62$               &$93.80\pm0.34$\\[3pt]
                        $\omega_{\rm NSS}$ (deg)            &$249.81\pm16.18$               &$199.57\pm5.01$                &$321.49\pm78.48$               &$103.41\pm12.72$           &$185.92\pm16.18$               &$51.91\pm30.56$                &$185.69\pm22.78$               &$92.95\pm1.58$\\[3pt]
                        lower $M_{\rm NSS}$ (\Mjup)&38.78                          &23.77                          &49                             &46.97                      &11.66                          &52.3                           &33.61                          &64.2\\[3pt]
                        $a_{0\rm{, NSS}}$ (mas)             &$0.37\pm0.02$                  &$0.91\pm0.02$                  &$1.22\pm0.14$                  &$0.98\pm0.03$              &$0.31\pm0.04$                  &$0.50\pm0.02$                  &$0.40\pm0.05$                  &$5.38\pm0.02$\\[3pt]
                        $K_{\rm NSS}$ (\mps)                &$1405.20\pm124.07$.            &$697.73\pm30.04$               &$527.12\pm641.58$.             &$1628.60\pm58.89$          &$460.15\pm86.89$               &$1268.67\pm69.51$              &$944.79\pm200.90$              &$2082.44\pm10.16$\\[3pt]
                    \hline
                        $P_{\rm b}$ (\days)                 &$281.89\pm0.09$                &$668.76_{-2.60}^{+2.54}$       &$718.67_{-6.35}^{+6.41}$       &$528.08\pm0.39$            &$588.58_{-2.50}^{+2.15}$       &$592.66_{-4.31}^{+4.92}$       &$558.67_{-2.47}^{+2.80}$       &$789.26_{-11.25}^{+10.08}$\\[3pt]
                        $K_{\rm b}$ (\mps)                  &$1233.31_{-3.91}^{+3.99}$      &$707.60_{-7.15}^{+7.78}$       &$335.09_{-3.89}^{+4.21}$       &$1418.62_{-1.76}^{+1.81}$  &$347.42_{-2.75}^{+2.15}$       &$1662.35_{-15.22}^{+17.60}$    &$738.92_{-1.87}^{+1.75}$       &$1984.95_{-18.19}^{+11.16}$\\[3pt]
                        $\lambda_0$ (deg)                   &$158.51_{-0.40}^{+0.41}$       &$-54.31_{-2.99}^{+2.91}$       &$152.79_{-6.05}^{+5.66}$       &$289.47_{-0.54}^{+0.52}$   &$452.61_{-5.31}^{+4.71}$       &$106.19_{-4.21}^{+4.67}$       &$425.03_{-3.63}^{+4.08}$       &$292.45_{-8.48}^{+7.20}$\\[3pt]
                        \sresinob                           &$-0.558\pm0.002$               &$-0.260\pm0.005$               &$0.838\pm0.005$                &$0.448\pm0.002$            &$-0.197\pm0.26$                &$-0.547\pm0.003$               &$0.341\pm0.004$                &$-0.575\pm0.006$\\[3pt]
                        \srecosob                           &$0.040\pm0.002$                &$-0.704\pm0.002$               &$0.192_{-0.029}^{+0.027}$      &$-0.245\pm0.002$           &$-0.528_{-0.013}^{0.015}$      &$-0.296\pm0.005$               &$-0.664\pm0.002$               &$0.060_{-0.005}^{+0.006}$\\[3pt]
                        $e_{\rm b}$                         &$0.313\pm0.002$                &$0.563\pm0.005$                &$0.739\pm0.006$                &$0.261\pm0.002$            &$0.318_{-0.012}^{+0.011}$      &$0.387\pm0.003$                &$0.557\pm0.002$                &$0.335\pm0.007$\\[3pt]
                        $\omega_{\rm b}$ (deg)              &$-85.94_{-0.23}^{+0.22}$       &$-159.72_{-0.37}^{+0.39}$      &$77.10_{-1.84}^{+1.98}$        &$118.68\pm0.28$            &$-159.52_{-2.84}^{+2.83}$      &$-118.43_{-0.49}^{+0.51}$      &$152.81\pm0.32$                &$-84.06_{-0.51}^{+0.60}$\\[3pt]
                        $M_{\rm b}\sin{i}$ (\Mjup)          &$30.32_{-0.91}^{+0.84}$        &$20.34\pm0.58$                 &$8.63\pm2.52$                  &$43.93\pm1.09$             &$10.30\pm0.45$                 &$59.52_{-2.38}^{+2.46}$        &$22.80_{-0.77}^{+0.75}$        &$63.35_{-2.03}^{+1.93}$\\[3pt]
                        $a_{\rm b}$ (au)                    &$0.75\pm0.01$                  &$1.34\pm0.02$                  &$1.47_{-0.23}^{+0.19}$         &$1.15\pm0.01$              &$1.20\pm0.02$                  &$1.34\pm0.03$                  &$1.27\pm0.02$                  &$1.44\pm0.02$\\[3pt]
                        $M_{b,i_{\rm NSS}}$ (\Mjup)         &$36.09^{+2.34}_{-2.38}$        &$24.55_{-1.01}^{+1.01}$        &$34.19\pm21.72$                &$44.29\pm1.11$             &$11.41\pm0.69$                 &$67.06^{+3.75}_{-3.68}$        &$28.81\pm5.30$                 &$63.49^{+1.93}_{-2.03}$\\[3pt]
                    \hline
                        $\gamma_{\rm HARPS-N}$ (\mps)       &$-510.58_{-1.70}^{+1.76}$      &$-280.73_{-6.22}^{+5.87}$      &$-118.56_{-2.24}^{+2.19}$      &$-706.32_{-1.13}^{+1.15}$  &$212.90_{-7.33}^{+8.04}$       &$137.47_{-18.61}^{+16.10}$     &$246.62_{-0.89}^{+0.98}$       &$-621.60_{-25.95}^{+16.97}$\\[3pt]
                        $j_{\rm HARPS-N}$ (\mps)            &$1.63_{-0.78}^{+0.75}$         &$3.78_{-0.40}^{+0.46}$         &$9.19_{-1.35}^{+1.70}$         &$2.98_{-0.86}^{+1.06}$     &$8.29_{-1.83}^{+2.62}$         &$12.19_{-1.73}^{+2.21}$        &$1.91_{-0.89}^{+1.08}$         &$6.80_{-0.98}^{+1.18}$\\[3pt]
                        $\gamma_{\rm HPF}$ (\mps)           &$-$                            &$-$                            &$-$                            &$-$                        &$-53.67_{-7.57}^{+7.84}$       &$-$                            &$-$                            &$-$\\[3pt]
                        $j_{\rm HPF}$ (\mps)                &$-$                            &$-$                            &$-$                            &$-$                        &$15.12_{-7.65}^{+10.14}$       &$-$                            &$-$                            &$-$\\[3pt]
                        $\gamma_{\rm NEID}$ (\mps)          &$-$                            &$-$                            &$-$                            &$-$                        &$-33.52_{-10.61}^{+11.07}$     &$-$                            &$-$                            &$-$\\[3pt]
                        $j_{\rm NEID}$ (\mps)               &$-$                            &$-$                            &$-$                            &$-$                        &$19.57_{-6.16}^{+9.49}$        &$-$                            &$-$                            &$-$\\[3pt]
                        $\gamma_{\rm FIES}$ (\mps)          &$-$                            &$-$                            &$-$                            &$-$                        &$61.13_{-9.94}^{+10.05}$       &$-$                            &$-$                            &$-$\\[3pt]
                        $j_{\rm FIES}$ (\mps)               &$-$                            &$-$                            &$-$                            &$-$                        &$10.40_{-7.25}^{+14.83}$       &$-$                            &$-$                            &$-$\\[3pt]
                    \hline
                \end{tabular}
            \end{sidewaystable*}

    \clearpage
    \section{Radial velocity and activity timeseries}
        \begin{figure}
            \includegraphics[width=\linewidth]{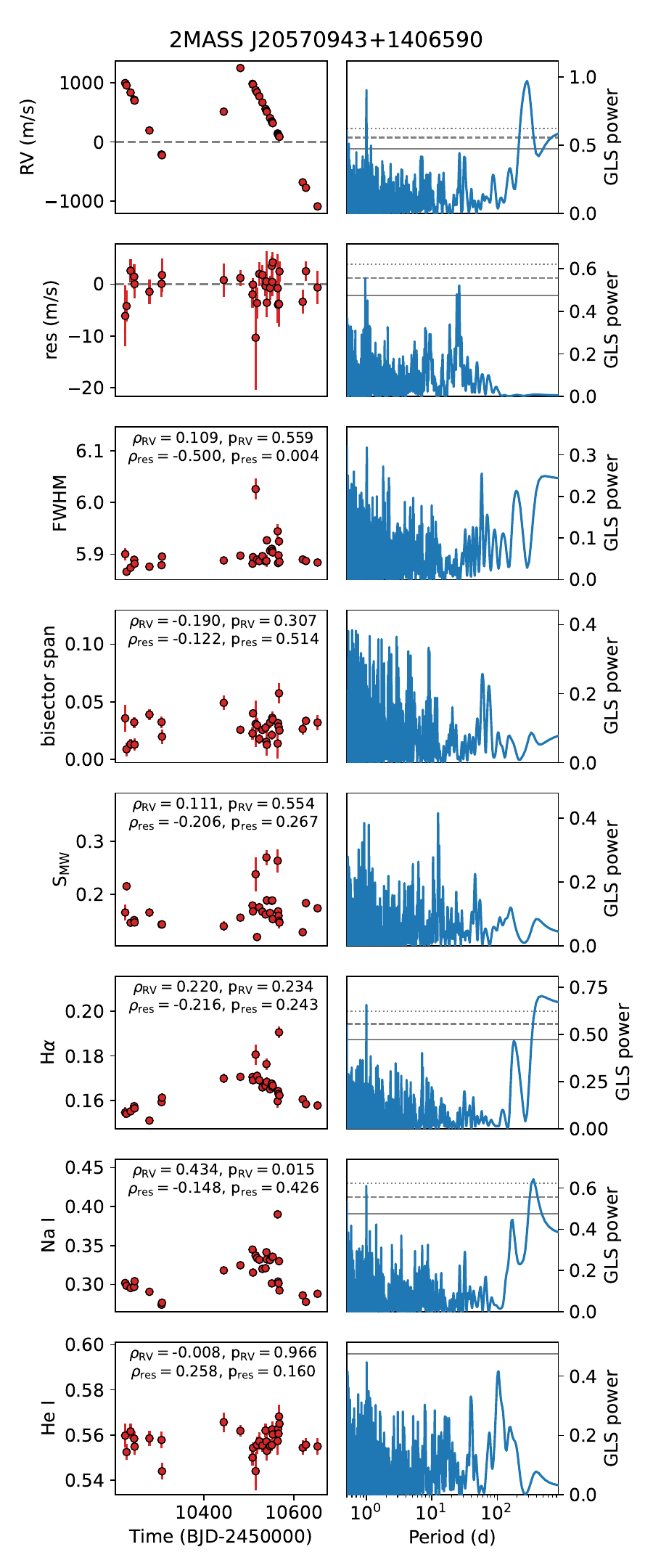}
            \caption{\textit{Left panels}: time series for the HARPS-N median-subtracted radial velocity data of 2MASS\,J20570943+1406590, post-Keplerian fit residuals and activity indices. For the activity indices time series, the Pearson correlation coefficient and its p-value with both original and residual RVs are noted. \textit{Right panels}: corresponding generalised Lomb-Scargle periodograms of the time series, with horizontal solid, dashed and dotted lines marking the 10\%, 1\% and 0.1\% FAP thresholds.}
            \label{fig:GFU24-timeseries}
        \end{figure}
        \begin{figure}
            \includegraphics[width=\linewidth]{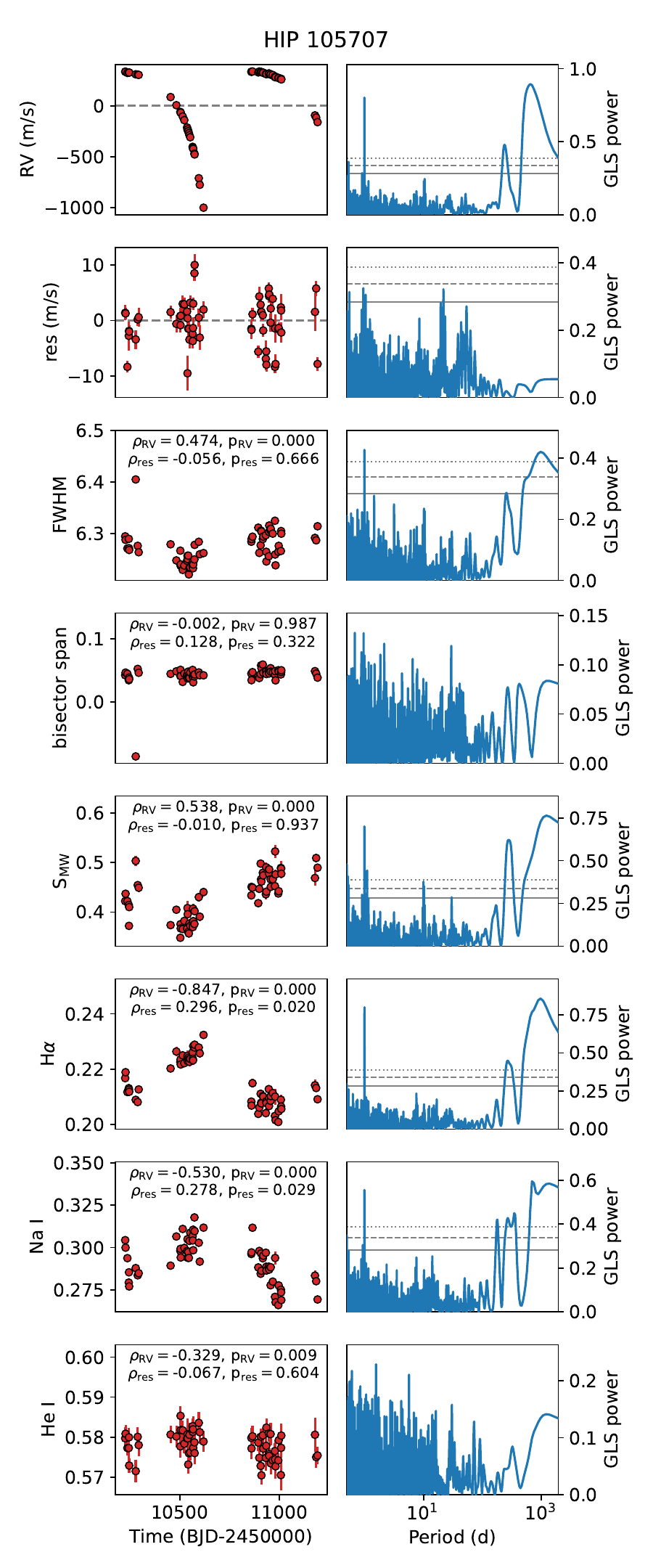}
            \caption{Same as Fig.~\ref{fig:GFU24-timeseries} but for star HIP\,105707}
            \label{fig:GFU25-timeseries}
        \end{figure}
        \begin{figure}
            \includegraphics[width=\linewidth]{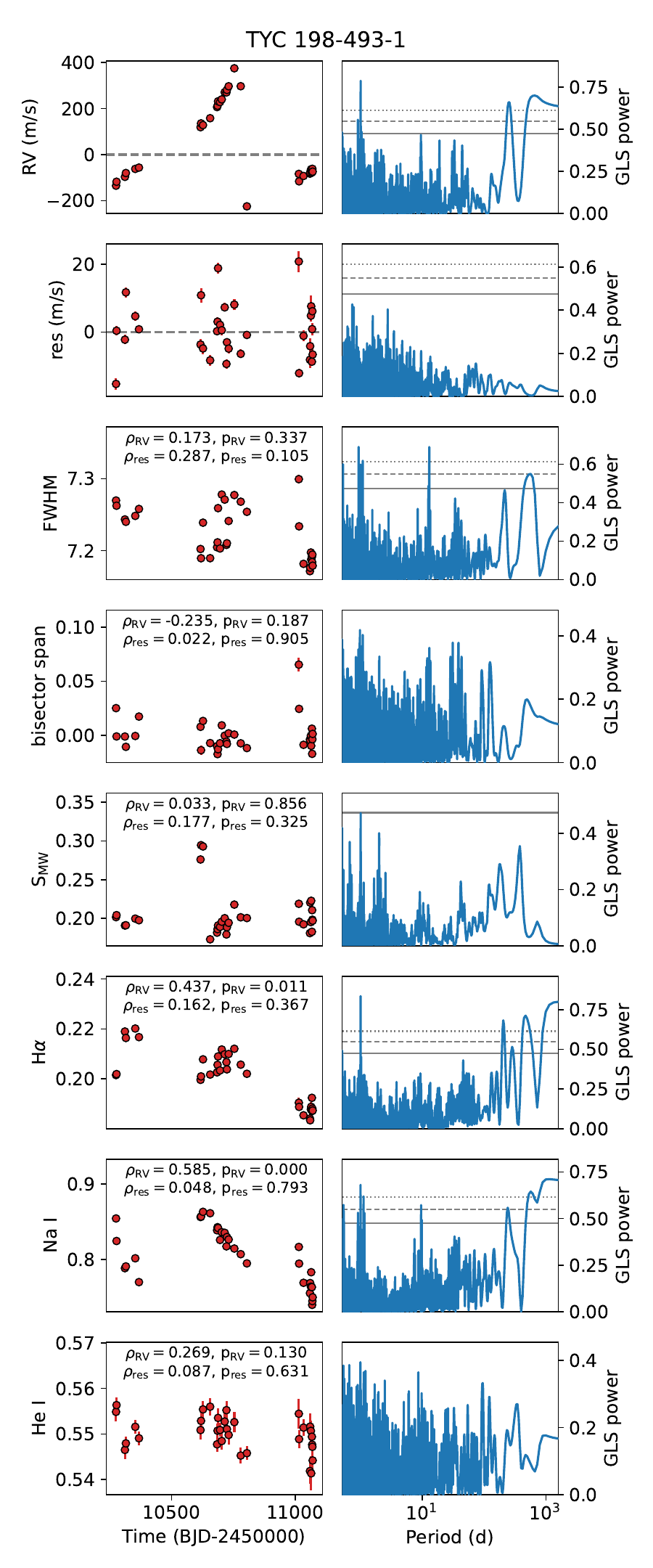}
            \caption{Same as Fig.~\ref{fig:GFU24-timeseries} but for star TYC\,198-493-1}
            \label{fig:GFU11-timeseries}
        \end{figure}
        \begin{figure}
            \includegraphics[width=\linewidth]{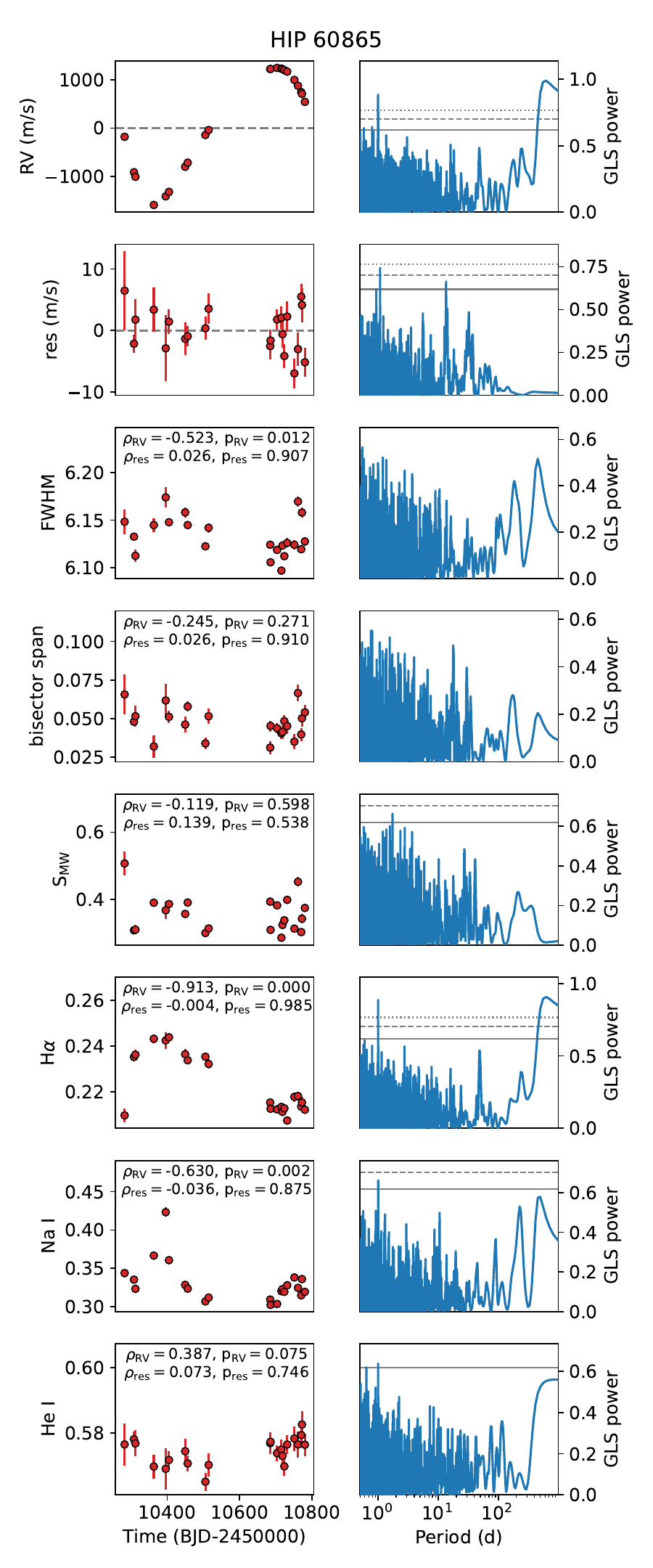}
            \caption{Same as Fig.~\ref{fig:GFU24-timeseries} but for star HIP\,60865}
            \label{fig:GFU16-timeseries}
        \end{figure}
        \begin{figure}
            \includegraphics[width=\linewidth]{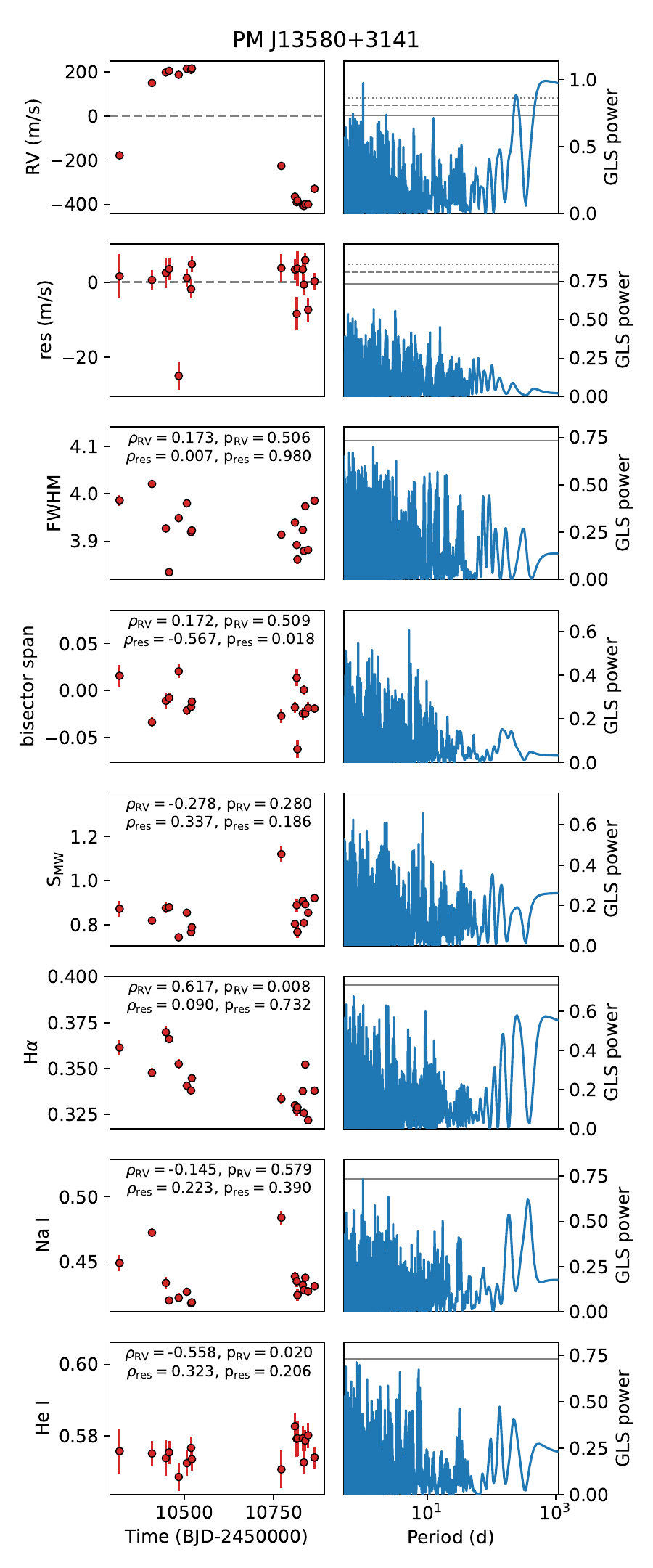}
            \caption{Same as Fig.~\ref{fig:GFU24-timeseries} but for star PM\,J13580+3141}
            \label{fig:GFU19-timeseries}
        \end{figure}
        \begin{figure}
            \includegraphics[width=\linewidth]{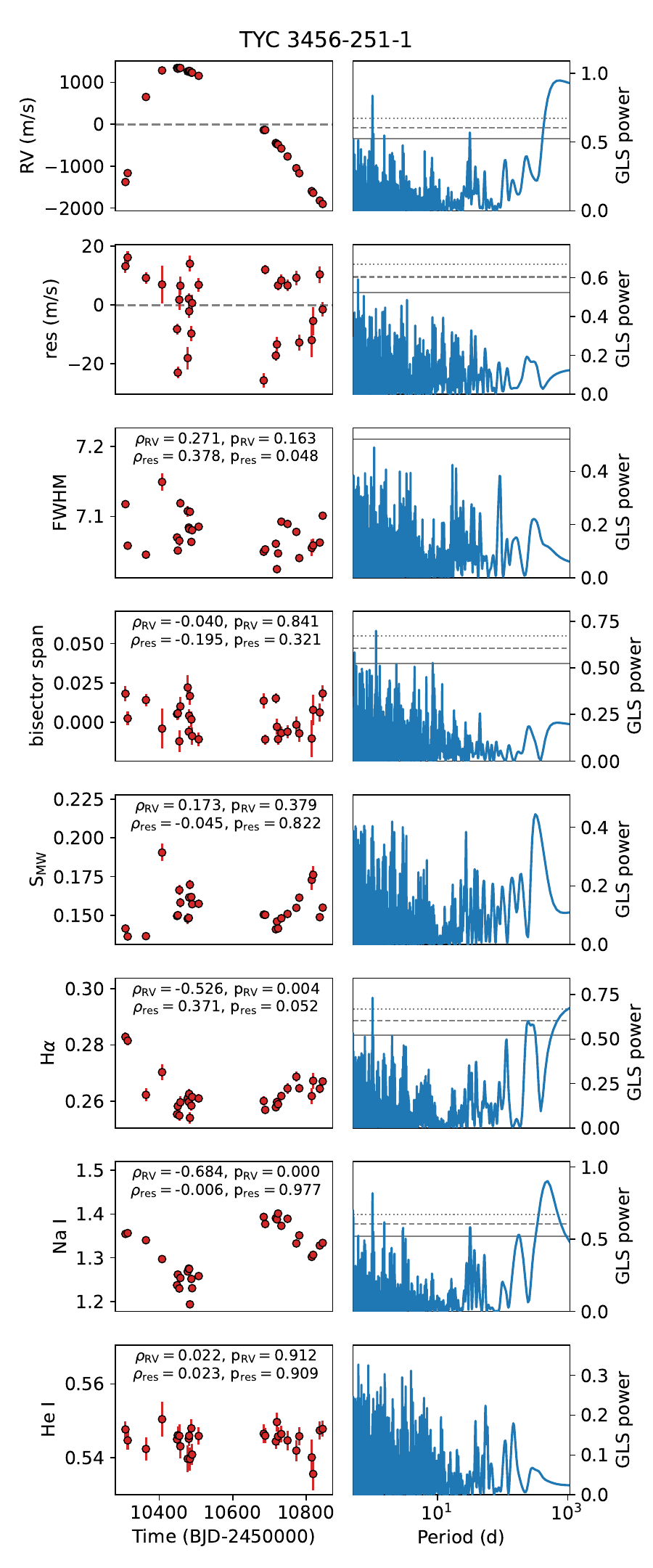}
            \caption{Same as Fig.~\ref{fig:GFU24-timeseries} but for star TYC\,3456-251-1}
            \label{fig:GFU17-timeseries}
        \end{figure}
        \begin{figure}
            \includegraphics[width=\linewidth]{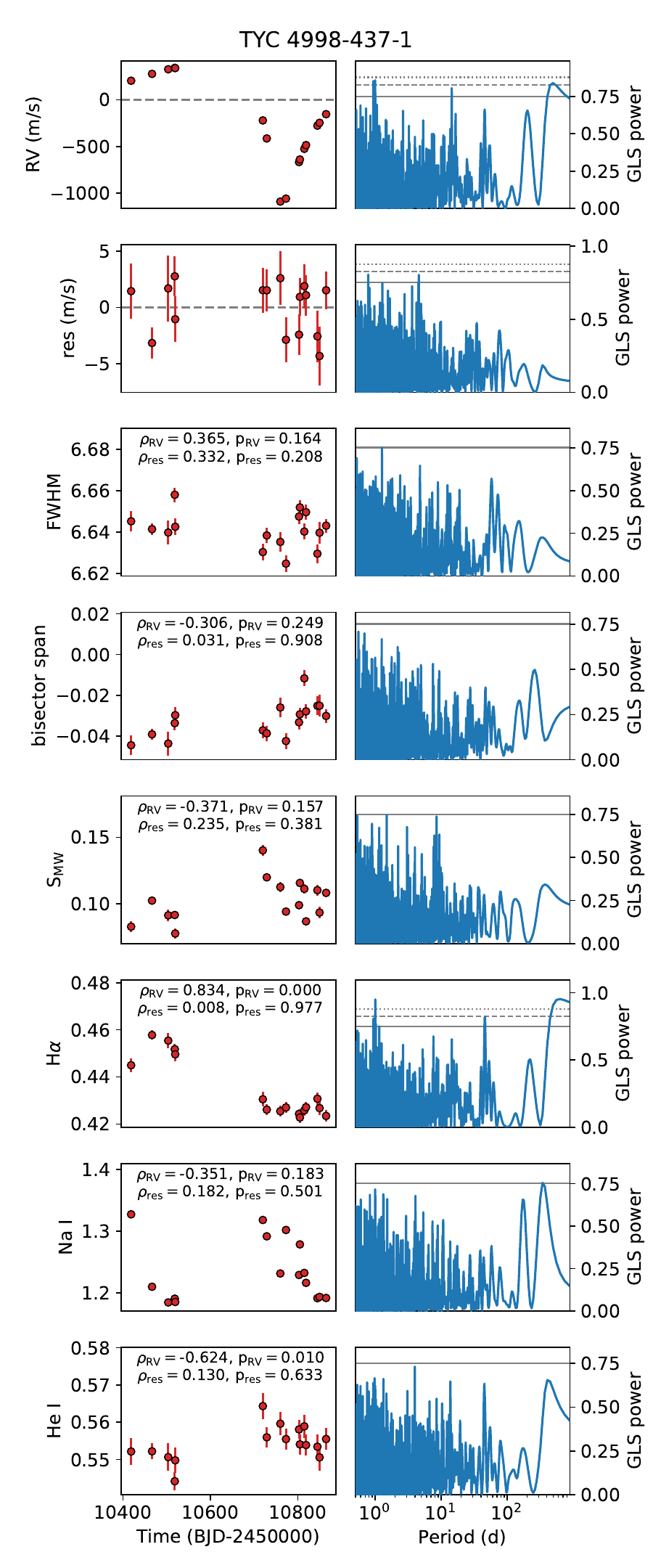}
            \caption{Same as Fig.~\ref{fig:GFU24-timeseries} but for star TYC\,4998-437-1}
            \label{fig:GFU20-timeseries}
        \end{figure}
        \begin{figure}
            \includegraphics[width=\linewidth]{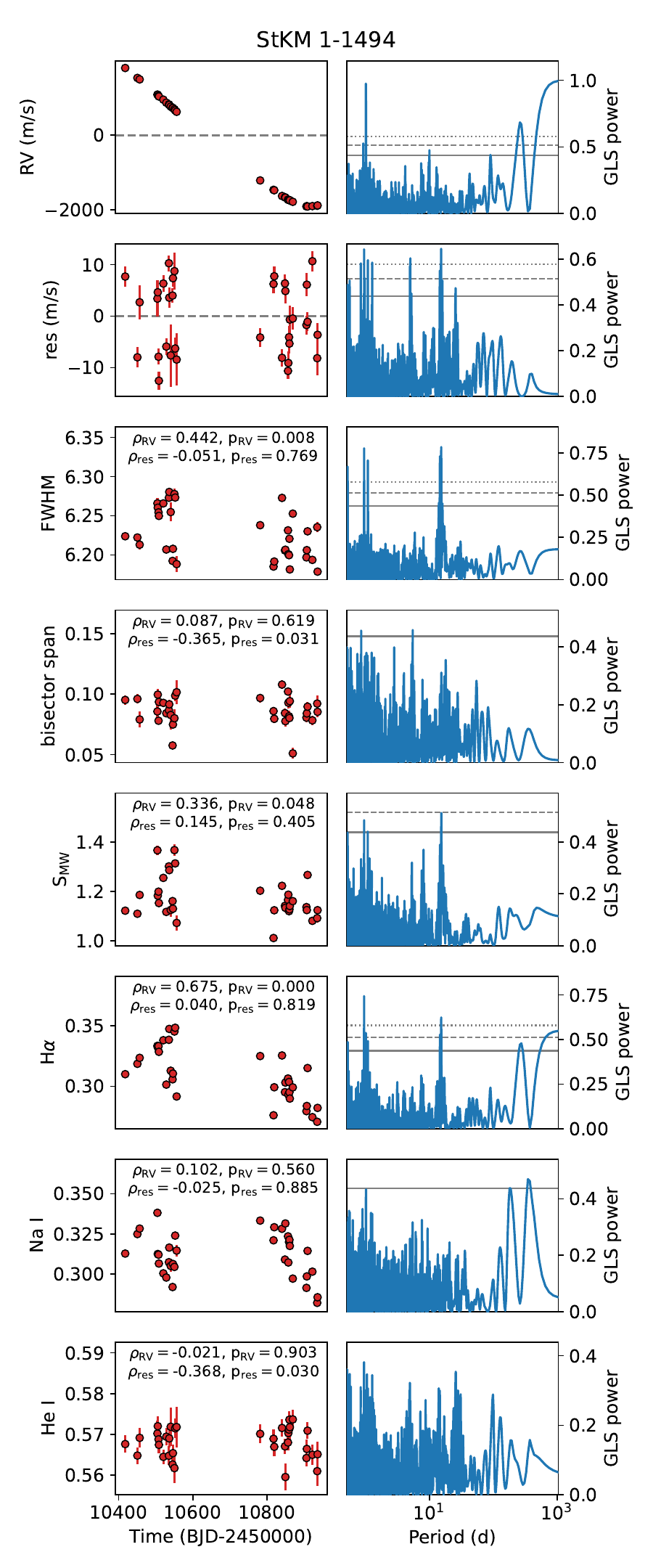}
            \caption{Same as Fig.~\ref{fig:GFU24-timeseries} but for star StKM\,1-1494}
            \label{fig:GFU22-timeseries}
        \end{figure}
\end{document}